\documentclass[fleqn,usenatbib]{mnras}

\usepackage{newtxtext,newtxmath}

\usepackage[T1]{fontenc}

\DeclareRobustCommand{\VAN}[3]{#2}
\let\VANthebibliography\thebibliography
\def\thebibliography{\DeclareRobustCommand{\VAN}[3]{##3}\VANthebibliography}

\usepackage{graphicx}	
\usepackage{amsmath}	
\usepackage{array}
\usepackage{ulem}

\newcommand\mysim{\mkern1.5mu{\sim}\mkern1.5mu}
\newcolumntype{P}[1]{>{\centering\arraybackslash}p{#1}}

\title[Magnification bias for complex sample selection functions]{Magnification bias in galaxy surveys with complex sample selection functions}

\author[M. von Wietersheim-Kramsta et al.]{
Maximilian von Wietersheim-Kramsta,$^{1}$
Benjamin Joachimi,$^{1}$
Jan Luca van den Busch,$^{2}$
\newauthor
Catherine Heymans,$^{2, 3}$
Hendrik Hildebrandt,$^{2}$,
Marika Asgari,$^{3}$
Tilman Tröster,$^{3}$
Sandra Unruh$^{4}$ and
\newauthor
Angus H. Wright$^{2}$
\\
$^{1}$Department of Physics and Astronomy, University College London, Gower Street, London, WC1E 6BT, UK \\
$^{2}$Ruhr University Bochum, Faculty of Physics and Astronomy, Astronomical Institute (AIRUB), German Centre for Cosmological Lensing, 44780 Bochum,\\ Germany \\
$^{3}$Institute for Astronomy, University of Edinburgh, Royal Observatory, Blackford Hill, Edinburgh EH9 3HJ, UK\\
$^{4}$Argelander-Institut für Astronomie, German Centre for Cosmological Lensing, Auf dem Hügel 71, D-53121 Bonn, Germany\\ 
}

\date{Accepted 2021 March 31. Received 2021 March 23; in original form 2021 January 13}

\pubyear{2021}

\begin{document}
\label{firstpage}
\pagerange{\pageref{firstpage}--\pageref{lastpage}}
\maketitle

\begin{abstract}
    Gravitational lensing magnification modifies the observed spatial distribution of galaxies and can severely bias cosmological probes of large-scale structure if not accurately modelled. Standard approaches to modelling this magnification bias may not be applicable in practice as many galaxy samples have complex, often implicit, selection functions. We propose and test a procedure to quantify the magnification bias induced in clustering and galaxy-galaxy lensing (GGL) signals in galaxy samples subject to a selection function beyond a simple flux limit. The method employs realistic mock data to calibrate an effective luminosity function slope, $\alpha_{\rm{obs}}$, from observed galaxy counts, which can then be used with the standard formalism. We demonstrate this method for two galaxy samples derived from the Baryon Oscillation Spectroscopic Survey (BOSS) in the redshift ranges $0.2 < z \leq 0.5$ and $0.5 < z \leq 0.75$, complemented by mock data built from the MICE2 simulation. We obtain $\alpha_{\rm{obs}} = 1.93 \pm 0.05$ and $\alpha_{\rm{obs}} = 2.62 \pm 0.28$ for the two BOSS samples. For BOSS-like lenses, we forecast a contribution of the magnification bias to the GGL signal between the multipole moments, $\ell$, of 100 and 4600 with a cumulative signal-to-noise ratio between 0.1 and 1.1 for sources from the Kilo-Degree Survey (KiDS), between 0.4 and 2.0 for sources from the Hyper Suprime-Cam survey (HSC), and between 0.3 and 2.8 for ESA Euclid-like source samples. These contributions are significant enough to require explicit modelling in future analyses of these and similar surveys. Our code is publicly available within the \textsc{MagBEt} module (\url{https://github.com/mwiet/MAGBET}).
\end{abstract}

\begin{keywords}
gravitational lensing: weak -- methods: data analysis -- methods: observational
\end{keywords}



\section{Introduction}
    Over the last few decades, weak gravitational lensing has become a powerful tool to directly measure the matter distribution of the late Universe, while allowing for the inference of the cosmological parameters which govern it. Surveys, such as the currently ongoing Kilo Degree Survey\footnote{\url{https://kids.strw.leidenuniv.nl}} (KiDS, \citealt{kuijken2015gravitational}), the Dark Energy Survey\footnote{\url{https://www.darkenergysurvey.org}} (DES, \citealt{flaugher2015dark}), the Hyper Suprime-Cam Subaru Strategic Program\footnote{\url{https://hsc.mtk.nao.ac.jp/ssp}} (HSC SSP, \citealt{aihara2018first}), have become increasingly limited by systematics rather than statistics as ever-growing sample sizes reduce uncertainties. The impact of the systematics will become even more exaggerated for the next generation of surveys, e.g. \textit{Euclid}\footnote{\url{https://www.euclid-ec.org}} \citep{laureijs2011euclid}, the Vera C. Rubin Observatory Legacy Survey of Space and Time\footnote{\url{https://www.lsst.org}} (LSST, \citealt{abell2009lsst}), and the Nancy Grace Roman Space Telescope\footnote{\url{https://roman.gsfc.nasa.gov}} (also known as WFIRST, \citealt{spergel2015wide}). For this reason, recent efforts have focused on improving our physical understanding of often neglected phenomena which can influence cosmological parameter inference based on shear and clustering measurements. These effects include intrinsic galaxy alignments \citep{kiessling2015galaxy, kirk2015galaxy, troxel2015intrinsic} and magnification \citep{hildebrandt2009cars,duncan2014complementarity,hildebrandt2016observational, unruh2020importance, thiele2020disentangling}. In this paper, we will focus on the magnification effects.\\
    
    While the magnification due to gravitational lensing partially manifests itself as a change in the angular diameter of an object, it also changes the observed solid angle of a field with respect to the intrinsic solid angle. This can affect the observed galaxy counts and their fluxes, leading to magnification effects, which have been detected in the past by \cite{chiu2016detection} and \cite{garcia2018weak}. It is important to note that this affects the counts of source galaxies and lens galaxies, such that the magnification due to large-scale structure can also change the shear-clustering cross-correlations (galaxy-galaxy lensing, GGL) and the clustering measurements \citep{hui2007anisotropic, ziour2008magnification,duncan2014complementarity, unruh2020importance, thiele2020disentangling}. Therefore, if this effect is not accurately modelled in such analyses, a \textit{magnification bias} can be induced. However, we also note that, in the literature and in this paper, the term \textit{magnification bias} is regularly used to refer magnification effects even when they are modelled.\\

    We break down the magnification effect into two separate phenomena: \textit{flux magnification} and \textit{lensing dilution}. The first is caused by an increase/decrease in the flux observed from a source due to gravitational lensing which can push otherwise unobserved galaxies over the flux limit or push galaxies with magnitudes below the flux limit out of the observational window. At the same time, lensing dilution increases/decreases the number of observed sources within a certain area of the sky by (de-)magnifying the solid angle behind the gravitational lens. The magnification effect can be measured directly from changes in the apparent size and magnitude of lensed galaxies \citep{schmidt2011detection} or by comparing the observed galaxy effective radii to the intrinsic radii derived from their surface brightness and stellar velocity dispersion \citep{huff2013magnificent}. Nonetheless, it is most commonly measured through the bias in the observed number density of sources \citep{scranton2005detection}. Since this bias directly contributes to the clustering and GGL signal, we will rely on this approach in our analysis.\\
    
    The constraining power of weak lensing samples is constantly growing \citep{troxel2015intrinsic, hikage2019cosmology, asgari2020kids} by including additional measurements \citep{abbott2018dark, abbott2019cosmological, abbott2019dark} and through joint analyses between different surveys like, for example, in the recent joint analysis of KiDS-1000 with BOSS (methodology described in \citealt{joachimi2020kids} and the results are shown in \citealt{heymans2020kids} and in \citealt{troster2020kids}). In all these analyses, the understanding of the systematics is be\-coming a priority. One potential systematic could appear from unaccounted magnification biases in the cluste\-ring signal of a non-flux-limited spectroscopic surveys such as BOSS \citep{dawson2012baryon} or DESI\footnote{\url{https://www.desi.lbl.gov}} \citep{aghamousa2016desi} or color-selected photometric samples such as DES \textsc{redMaGiC} \citep{rozo2016redmagic} or luminous red galaxy (LRG) samples \citep{vakili2020clustering}. Thus also biasing the GGL correlations with shear signal from weak lensing surveys.\\
    
    This paper aims to provide a method for estimation of the magnification bias for surveys which have complex sample selection functions which are not purely flux/magnitude-limited. We use the standard framework for estimating the magnification bias from observables in flux-limited surveys as a basis for the parametrisation of a semi-empirical model for non-flux-limited surveys. This model is then tested by comparing the estimates for the magnification bias in BOSS observations \citep{dawson2012baryon} to the estimates from MICE2 cosmological simulations. We then use our results to forecast some of the potential biases which could be induced in a joint analysis of KiDS-1000 or HSC Wide with BOSS and a \textit{Euclid}-like survey with a DESI-like survey.\\
    
    This article is structured in the following manner. In section~\ref{background}, the theoretical background is described. In section~\ref{methodology}, we provide an outline and presentation of our methods and simulations. The magnification bias estimates from a BOSS-like galaxy population are presented in section~\ref{non-flux-lim}. The forecasts for current and future joint analyses are found in section~\ref{forecasts}. Lastly, we conclude the paper and provide an outlook in section~\ref{conclusion}. Appendix~\ref{flux-lim} repeats the analysis shown in section~\ref{non-flux-lim} for a magnitude limited galaxy sample.

\section{Theoretical background}\label{background}

\subsection{Magnification bias for flux-limited surveys}\label{back:mag}
    
    As described in the review by \citeauthor{bartelmann2001weak} (\citeyear{bartelmann2001weak}), a lensed population of galaxies with a cumulative galaxy count $N$ at redshift $z$, given a flux limit of $S$, can be described in terms of the unlensed population, $N_{0}$, as
    \begin{equation}
        N(> S, z) =  
            \frac{1}{\mu(z)} N_{0}\Bigg(>\frac{S}{\mu(z)}, z\Bigg) \, ,
        \label{eqn:n}
    \end{equation}
    \noindent
    where $\mu(z)$ is the magnification for a redshift $z$. Here, the $1 / \mu(z)$ factor accounts for the dilution of galaxies due to magnification. The unlensed population has been observationally shown to be similar to a power law in flux (in particular, for faint galaxies) given by
    \begin{equation}
        N_{0}(>S, z) = 
            A 
            S^{-\alpha}
            p_{0}(z; S) \, ,
        \label{eqn:n0}
    \end{equation}
    \noindent where $A$ and $\alpha$ parametrise the power law and $p_{0}(z; S)$ is the redshift probability distribution of the galaxies. Taking the ratio of these two populations, assuming that we can approximate the $\mu(z)$ with the magnification $\mu$ of a fiducial source at infinity (which should hold mainly at low redshifts, \citeauthor{bartelmann2001weak}, \citeyear{bartelmann2001weak}) and integrating over redshift, we get the following expression:
    
    \begin{equation}
        \frac{N(>S)}{N_{0}(>S)} = 
            \mu^{\alpha-1}.
        \label{eqn:ratio}
    \end{equation}
    
    \indent If $\alpha \approx 1$, we can see from equation~(\ref{eqn:ratio}) that the magnification bias would vanish (with slight deviations from this depending on the redshift range). The magnification can be related directly to the local surface density $\kappa$ in the weak len\-sing limit ($|\kappa| \ll 1$, $|\gamma| \ll 1$) with $\mu \approx 1 + 2\kappa$ \citep{broadhurst1995gravitational}. Therefore, one can relate $\kappa$ to the relative difference between the magnified and the unmagnified galaxy populations and the exponent $\alpha$ of the flux power spectrum with
    
    \begin{equation}
    	\frac{N(>S) - N_{0}(>S)}{N_{0}(>S)}  \approx 
    	    2(\alpha_{\kappa}-1) 
    	    \kappa,
    	\label{eqn:reldiff}
    \end{equation}
    \noindent where $\alpha_{\kappa}$ is the same as the $\alpha$ in equation~(\ref{eqn:ratio}) in the weak lensing limit. When analysing samples with a complex selection function, equation~(\ref{eqn:reldiff}) does not necessarily apply anymore. Nonetheless, we use the parameter $\alpha_{\kappa}$ as an analogue to estimate the magnitude of the magnification bias in a given galaxy sample.
    
    \subsection{Estimating the magnification bias in flux-limited surveys} \label{back:obs}

    By considering equations~(\ref{eqn:n}), (\ref{eqn:n0}), and the definition of magnitude as a function of flux, one can derive that $\alpha_{\rm{obs}}$ can be determined from the differential galaxy count $n(m)$ over a given band magnitude range from $m$ to $m+dm$ as follows \citep{binggeli1988luminosity, bartelmann2001weak, hildebrandt2009cars},
    \begin{equation}
    	\alpha_{\rm{obs}}(m) = 
        	2.5 \frac{\rm{d} 
        	\log_{10}[\mathit{n}(\mathit{m})]}{\rm{d}\mathit{m}}.
    	\label{eqn:alpha_obs}
    \end{equation}
    One could get the same estimates of $\alpha_{\rm{obs}}(\mathit{m})$ (at least, for a flux-limited sample) by replacing $n(m)$ in equation~(\ref{eqn:alpha_obs}) with the cumulative galaxy count distribution. However, here we choose to derive $\alpha_{\rm{obs}}(\mathit{m})$ from the differential distribution, $n(m)$, instead, because we find that it gives more robust estimates when deviating from the flux-limited case. Also, note that sometimes the differential galaxy count distribution is given over flux, $S$, instead of magnitude, $m$. Then, $\alpha_{\rm{obs}}(S) + 1$ is given by $-\rm{d} \rm{log_{10}}[\mathit{n(S)}]/\rm{d}\rm{log_{10}}(\mathit{S})$. \\ 
    
    This $\alpha_{\rm{obs}}$ near the faint end of the galaxy population is considered as an \textit{effective} luminosity function slope if it is consistent with the $\alpha_{\kappa}$ value given by equation~(\ref{eqn:reldiff}). Therefore, by estimating the luminosity function slope, $\alpha_{\kappa}$, through the observed $\alpha$, one can estimate the systematic effects that may be introduced to galaxy number counts through the magnification bias, and therefore the systematics affecting the clustering and GGL signals derived from this observable.
    
    \subsection{Signal modelling} \label{back:signal}
    In accordance with the framework outlined in Section 2 of \citet{joachimi2020kids} as the methodology for the inference of cosmolo\-gical parameters from KiDS-1000, we opt to quantify the influence of the magnification bias on cosmology through its contribution to the GGL angular power spectra. These angular power spectra are line-of-sight projections of the three-dimensional matter power spectrum. We express the observable GGL angular power spectrum correlating galaxy positions, $\rm{n}$, and galaxy shapes, $\epsilon$, as a linear functional of derived statistics as
    \begin{equation}
        	C_{\rm{n}\epsilon}^{(ij)}(\ell) = 	C_{\rm{gG}}^{(ij)}(\ell) + 	C_{\rm{gI}}^{(ij)}(\ell) + C_{\rm{mG}}^{(ij)}(\ell),
        	\label{eq:power_spectra}
    \end{equation}
    \noindent where $i$ is the index for lens galaxy redshift bins, $j$ is the index of the source galaxy samples, gG stands for the cross-correlation between the lens galaxy distribution and the source gravitational shear, gI stands for the intrisinc alignment of source galaxies physically close to foreground lenses and mG stands for the correlation between gravitational shear and the lensing-induced magniﬁcation bias in the lens sample. $C_{\rm{ga}}^{(ij)}(\ell)$ for $a \in \{G, I\}$ are defined as Limber-approximated line-of-sight projections of the three-dimensional cross-power spectrum between the galaxy and matter distribution, $P_{\rm{gm}}$, given by \citep{kaiser1992weak, loverde2008extended}
    
    \begin{equation}
        C_{\rm{ga}}^{(i j)}(\ell) = 
        \int^{\chi_{\rm{hor}}}_{0} 
            \frac{n_{\rm{L}}^{(i)}(\chi) W_{\rm{a}}^{(j)}(\chi)}{f^{2}_{K}(\chi)}
            P_{\rm{gm}}\Bigg(\frac{\ell + 1/2}{f_{\rm{K}}(\chi)}, \chi \Bigg) \, ,
            \label{eqn:aps_ga}
    \end{equation}
    
    \noindent where $\chi$ is the comoving distance, $\chi_{\rm{hor}}$ is the comoving distance to the horizon, $n_{\rm{L}}^{(i)}$ is the comoving distance distribution of the lens sample $i$ and $f_{\rm{K}}$ is the comoving angular diameter distance.  $W_{\rm{G}}^{(j)}$ is the weak lensing kernel and is given by
    
    \begin{equation}
        W_{\rm{G}}^{(j)}(\chi) = 
        \frac{3 H_{\rm{0}}^{2} \Omega_{\rm{m}}}{2 c^{2}}
        \frac{f_{\rm{K}}(\chi)}{a(\chi)}
        \int^{\chi_{\rm{hor}}}_{\chi} \rm{d}\chi'
        n_{\rm{S}}^{(j)}(\chi')
        \frac{\mathit{f}_{\rm{K}}(\chi' - \chi)}{\mathit{f}_{\rm{K}}(\chi')},
    \end{equation}
    
    \noindent where $H_{\rm{0}}$ is the Hubble constant, $\Omega_{\rm{m}}$ is the matter density parameter, $c$ is the speed of light, $a$ is the scale factor and $n_{\rm{S}}^{(j)}$ is the comoving distance distribution of the source sample $j$. $W_{\rm{I}}^{(j)}$ is the intrinsic alignment (IA) kernel. Here, we choose an IA kernel in accordance with the non-linear alignment model (NLA, \citealt{bridle2007dark}) given by
    
    \begin{equation}
        W_{\rm{I}}^{(j)}(\chi) =
        - A_{\rm{IA}}
        \frac{C_{1} \rho_{\rm{cr}} \, \Omega_{\rm{m}}}{D(a[\chi])}
        \, n_{\rm{S}}^{(j)}(\chi),
    \end{equation}
    
    \noindent where $A_{\rm{IA}}$ is the IA amplitude, $z_{\rm{pivot}}$ is an arbitrary pivot which is set to 0.3 in line with previous IA analyses \citep{joachimi2011constraints}, $C_{1}$ denotes a normalisation constant, $\rho_{\rm{cr}}$ is the critical density, $D$ is the linear growth factor normalized to unity at the present day. We normalise the IA kernel by setting $C_{1}\rho_{\rm{cr}} \approx 0.0134$, i.e. $C_{1} = 5 \times 10^{-14} (h^{2} M_{\odot} /Mpc^{-3})^{-2}$, in accordance with the value from \citealt{hirata2004intrinsic} and \citealt{bridle2007dark} which is set using the galaxy ellipticity measurements from SuperCOSMOS \citep{hambly2001supercosmos, brown2002measurement}.\\
    
   The magnification term in equation~(\ref{eq:power_spectra}) is modelled as
    \begin{equation}
    	C_{\rm{mG}}^{(ij)}(\ell) = 2 (\alpha_{\rm{obs}}^{(i)} - 1) C_{\rm{GG}}^{(ij)}(\ell),
    	\label{eqn:ps_gm}
    \end{equation}
    \noindent where $i$ again indexes lens galaxy samples, $j$ indexes source samples, mG stands for  the lensing-induced magniﬁcation bias in the lens sample and GG stands for shear-shear correlation signal. $ C_{\rm{GG}}^{(ij)}(\ell)$ is defined as the cosmic shear angular power spectrum purely from gravitational lensing effects, i.e. without any intrinsic alignment signals, and is given by
    
    \begin{equation}
        C_{\rm{GG}}^{(i j)}(\ell) = 
        \int^{\chi_{\rm{hor}}}_{0} 
            \frac{W_{\rm{G}}^{(i)}(\chi) W_{\rm{G}}^{(j)}(\chi)}{f^{2}_{K}(\chi)}
            P_{\rm{m,nl}}\Bigg(\frac{\ell + 1/2}{f_{\rm{K}}(\chi)}, z(\chi) \Bigg) \, ,
    \end{equation}
    \noindent where $P_{\rm{m,nl}}$ is the non-linear matter power spectrum. This power spectrum is computed with a non-perturbative model using \textsc{HMCode} \citep{mead2015accurate, mead2016accurate} integrated within \textsc{CAMB}\footnote{Code for Anisotropies in the Microwave Background; \url{https://camb.info}} \citep{lewis2000efficient, lewis2002cosmological, howlett2012cmb}. \textsc{HMCode} incorporates baryonic feedback in its halo modelling approach. We solely parametrise the baryonic feedback model using one free parameter, $A_{\rm{bary}}$, in line with \cite{hildebrandt2017kids}. The non-linear matter power spectrum $P_{\rm{m,nl}}$ is also used to compute the cross-power spectrum between the galaxy and matter distribution $P_{\rm{gm}}$ used in equation~(\ref{eqn:aps_ga}) as in the analysis shown in \cite{joachimi2020kids}.

\section{Methodology} \label{methodology}
The method outlined in this paper aims to provide an accurate estimate of the \textit{effective} luminosity function slope, $\alpha$, of a galaxy sample with a complex sample selection. This estimate can be used to quantify the magnification bias in clustering and GGL lensing analyses. To achieve this, we rely on realistic weak lensing simulations to ca\-librate the $\alpha_{\rm{obs}}$ estimate from observables, based on equation~(\ref{eqn:alpha_obs}), such that it agrees with the value of $\alpha_{\kappa}$ derived from unobservable quantities using equation~(\ref{eqn:reldiff}). The procedure gives a magnitude range that yields the most optimal $\alpha_{\rm{obs}}$ value. This value is used to estimate $\alpha_{\rm{obs}}$ from observations. If the simulations are accurate, $\alpha_{\rm{obs}}$ should agree with the underlying $\alpha_{\kappa}$ even though it cannot be directly measured.

\subsection{BOSS DR12 data}\label{method:boss_data}
    We develop our method using lens samples derived from the Sloan Digital Sky Survey (SDSS)-III BOSS \citep{eisenstein2011sdss, dawson2012baryon}. BOSS is a spectroscopic survey with a complex sample selection function which is commonly used in cosmological analyses of galaxy clustering and GGL \citep{ alam2017clustering, sanchez2017clustering, beutler2017clustering, troster2019cosmology, speagle2019galaxy, heymans2020kids}. For more details about the nature of the galaxy selection process, see \cite{alam2015eleventh}. A lens galaxy sample selected in such a way could be introducing a substantial magnification bias in any analysis, while its complexity does not allow to measure it with current means. For the BOSS sample, the bias becomes even more important to model, because it is commonly used in GGL analysis with the source galaxy samples of weak lensing surveys whose footprint significantly overlaps with the BOSS footprint.\\
    
\begin{figure}
       \centering
       \includegraphics[width=\columnwidth]{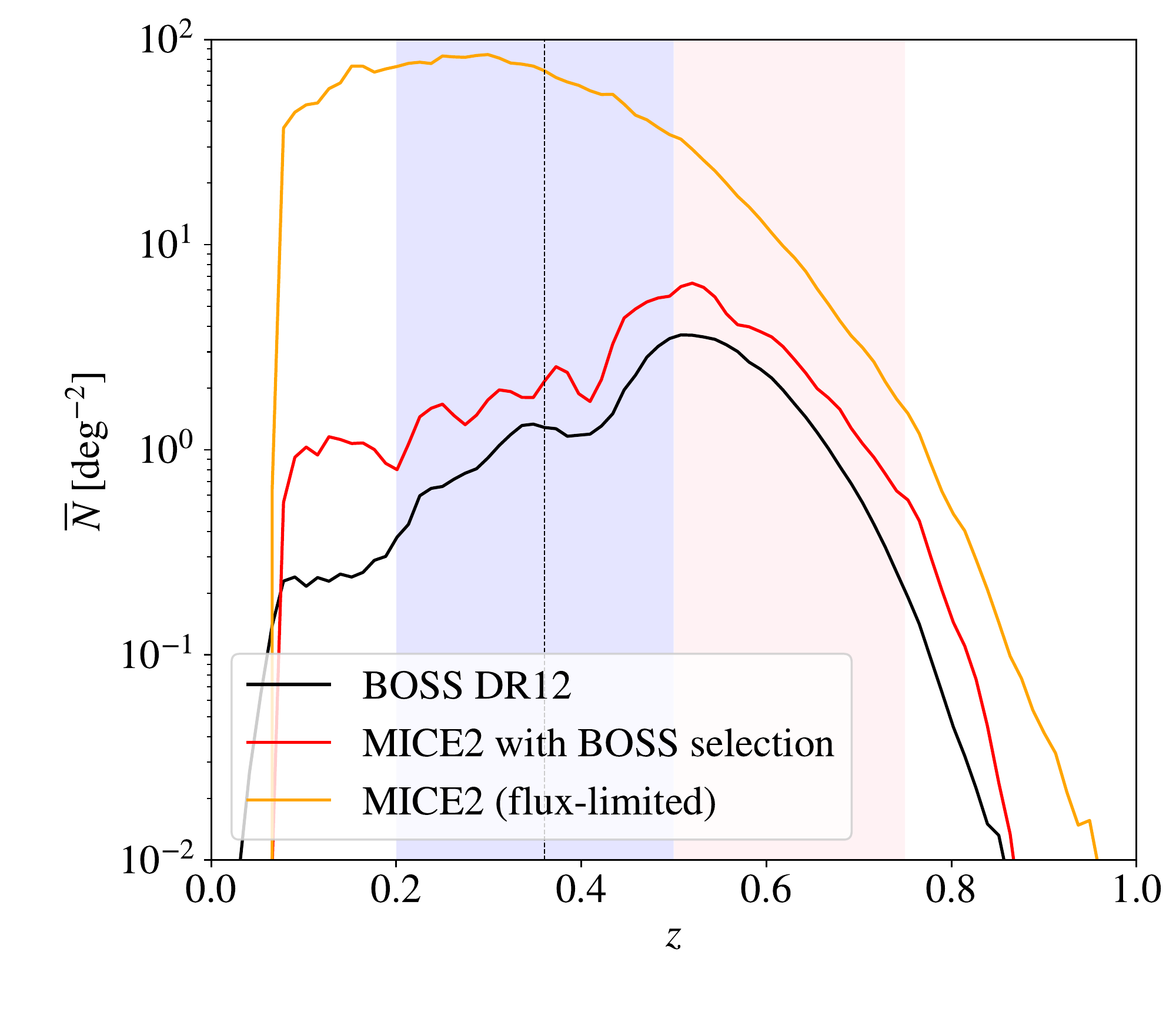}
          \caption{Galaxy counts per unit area on the sky, $\overline{N}$, for 100 redshift bins within $0  < z \leq 1$. The SDSS DR12 BOSS sample is shown in black, the MICE2 sample with the BOSS selection function in red and the flux-limited MICE2 sample in orange. The blue area marks the domain between $z = 0.2$ and $z \leq 0.5$ which defines the zlow bin, while the red area marks the domain of the zhigh bin ($0.5 < z \leq 0.75$). The dashed black horizontal line indicates the boundary between the LOWZ and the CMASS samples within the BOSS DR12 sample at $z \sim 0.36$.}
         \label{fig:counts}
   \end{figure}

    For this work, we use the photometric data from the final data release of BOSS, DR12 \citep{alam2015eleventh} with the same target selection as in \cite{sanchez2017clustering}. This sample combines the BOSS LOWZ and CMASS galaxy samples to produce a catalogue which covers approximately 9300 deg$^2$ \citep{reid2016sdss}. Its normalised redshift distribution can be seen in figure~\ref{fig:counts}. The sample is then split into two redshift ranges: "zlow" ($0.2 < z \leq 0.5$) and "zhigh" ($0.5 < z \leq 0.75$). From this photometric data, we use SDSS composite model (cmodel) band magnitudes which are defined in \cite{stoughton2002sloan}.
        
    \begin{figure}
       \centering
       \includegraphics[width=\hsize]{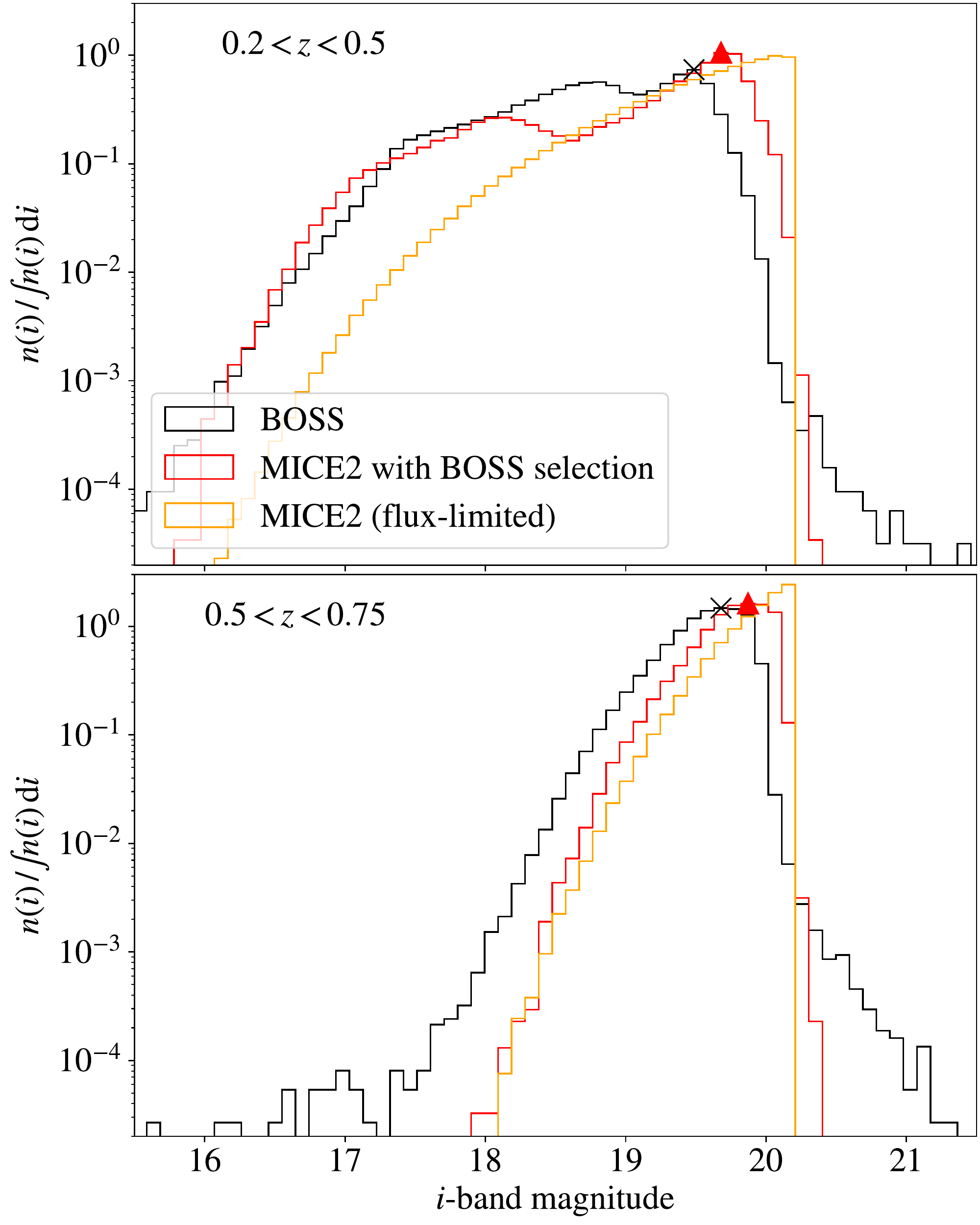}
          \caption{Normalised differential galaxy count distribution, $n(i)$, with respect to the $i$-band magnitude. The BOSS sample is shown in black, the MICE2 sample in red, while the flux-limited MICE2 sample is shown in yellow. In the top plot, we see the population of galaxies with $0.2 < z \leq 0.5$ and at the bottom, the population of galaxies with $0.5 < z \leq 0.75$. The black cross indicates the effective magnitude limit determined for the BOSS sample by finding the faintest prominent peak in $n(i)$. The red triangle indicates the same for the MICE2 mock sample.}
         \label{fig:count_hist}
   \end{figure}

\subsection{MICE2 simulations}\label{method:data}
    For the analysis discussed in section~\ref{non-flux-lim} and in appendix~\ref{flux-lim}, we rely on datasets of simulated galaxies, selected from the MICE2 galaxy mock catalogue \citep{Fosalba15a,Fosalba15b,Carretero15,Crocce15,Hoffmann15}. This catalogue is based on the MICE dark matter-only simulation, generated from $7 \times 10^{10}$ particles in a box with a side length of 3 Gpc and assuming a $\Lambda$CDM cosmological model with $\Omega_{\rm m} = 0.25$, $\Omega_\Lambda = 0.75$, $\Omega_{\rm b} = 0.044$ and $h=0.7$. A light cone, spanning $\mysim 5000$ deg$^{2}$, is constructed from this simulation box and populated with galaxies up to a redshift of $z=1.4$ using a hybrid Halo Occupation Distribution (HOD) and Halo Abundance Matching (HAM) technique. Additionally, MICE2 embeds gravitational lensing by providing estimates of the shear components, convergence as well as true and lensed position for each galaxy. MICE2 derives weak lensing properties by constructing all-sky shells in steps of 35 Myr of lookback time \citep{Fosalba15b}. These are then projected into \textsc{HealPix} maps \citep{gorski2005healpix} from which the convergence is computed using the Born-approximation \citep{Fosalba15b}. Therefore, galaxies within the same \textsc{HealPix} pixel inherit the same lensing properties which are, due to this limitation, accurate down to scales of ~1 arcmin. We compute the magnified galaxy magnitudes according to equation~(\ref{eqn:magn_mag}) which uses the weak lensing assumption $|\kappa| \ll 1$ by approximating $\mu \approx 1 + 2\kappa$.\\
    
    We start from this MICE2 input catalogue and apply an evolutionary correction to the provided SDSS $g^\prime r^\prime i^\prime$-band magnitudes and calculate an additional set of magnitudes
    
    \begin{equation}
        m^{\rm mag} = m^{\rm evo} - 2.5 \log_{10}(1 + 2\kappa) \, ,
        \label{eqn:magn_mag}
    \end{equation}
    \noindent that factor in magnification, where $m^{\rm evo}$ are the evolution corrected MICE2 magnitudes and $\kappa$ the convergence (see \citealt{vandenBusch20} for details). This allows us later to separate the effects of lensing dilution and magnification in the mock data. \citealt{unruh2020importance} recently showed that the weak lensing assumption ($|\kappa|\ll 1, |\gamma|\ll1$) used to derive equation~(\ref{eqn:reldiff}) and equation~(\ref{eqn:magn_mag}) might lead to biases when simulating magnified galaxy samples. Since $99.9\%$ of the galaxies in the MICE2 simulations have $|\kappa| < 0.09$, the assumption should still hold. However, it should be investigated in the future whether this is really the case.\\
    
    Finally, we select two samples from this base catalogue, one with an arbitrary magnitude limit in the SDSS $i$-band at $m^{\rm mag}_i \leq 20.2$ (applied in appendix~\ref{flux-lim}) and one that resembles the SDSS BOSS survey, using a target selection similar to \citet{eisenstein2011sdss} (applied in section~\ref{non-flux-lim}). The details of this BOSS mock sample selection are summarised in \citet{vandenBusch20}.\\
    
    The $i$-band number counts of these two samples and the original BOSS data is shown in figure~\ref{fig:count_hist}. In figure~\ref{fig:count_hist}, it becomes apparent how the BOSS selection function differs from a flux-limited sample. The cut-off of the galaxy population at the magnitude limit is not as pronounced, while the $n(i)$ no longer increases monotonically, especially at low redshifts. The galaxy counts per unit area as a function of redshift of the three samples is shown in figure~\ref{fig:counts}. Here we see how with the BOSS selection function applied, the redshift distribution is altered in a highly non-linear manner causing it to be multi-peaked with a main peak at $z \sim 0.5$. The magnitude-limited sample, on the other hand, follows a roughly single-peaked distribution dominated by low redshift galaxies ($z \sim 0.3$). \\

     Having knowledge of the underlying matter distribution allows us to compare estimates of the scale of flux magnification through $\alpha_{\rm{obs}}(m)$ from observables as given by equation~(\ref{eqn:alpha_obs}) with the $\alpha_{\kappa}$ estimate as given by equation~(\ref{eqn:reldiff}). When analysing the MICE2 mock observations, we only consider the SDSS model $i$-band magnitude, due to a lack of available SDSS cmodel magnitudes from the simulations.\\

     As a sanity check of our methods outlined in section~\ref{method:sim}, we conduct an estimate of the magnification bias induced by a flux/magnitude-limited sample selection function on a galaxy survey over an eighth of the sky in appendix~\ref{flux-lim}. For this, we use the MICE2 simulations to obtain the positions and magnitudes of galaxies before and after magnification, while knowing the true underlying matter density. We set the magnitude limit in the $i$-band to a magnitude of 20.2 (similar to the magnitude limit in the $i$-band of the BOSS survey). We find that the calibrated $\alpha_{\rm{obs}}$ values accurately recover $\alpha_{\kappa}$ near the faint limit. At the same time, the $\alpha_{\rm{obs}}$ estimates are robust over large changes in the calibration range chosen, showing that the power law approximation holds within $\mysim 1 \sigma$ over $\Delta i \sim 1$ and within $\mysim2 \sigma$ over the whole magnitude range for both \textit{zlow} and \textit{zhigh}.\\

    To conduct the analysis for the case where the target selection function is not flux or magnitude limited, we select a $\mysim5000$ deg$^2$ area from the MICE2 simulations and apply the aforementioned sample selection function to it. The $i$-band magnitude distribution of the BOSS and MICE2 galaxies within each of the redshift bins is shown in figure~\ref{fig:count_hist}. Here, we see that, although the overall shape of the population is similar between the BOSS and the MICE2 galaxies, the MICE2 objects are consistently shifted towards the fainter end of the distribution. This is at least partially caused by the fact that the BOSS magnitudes are $i$-band cmodel magnitudes and the MICE2 magnitudes are SDSS model $i$-band magnitudes. In addition, the MICE2 simulations with a BOSS-like selection function do not seem to capture the population of galaxies at the extremes of the magnitude distribution. Both of these biases might also be due to some assumptions in the galaxy formation and evolution models used in the MICE2 simulations. In addition, the fiducial cosmology assumed for the simulations might not agree with the cosmological parameters preferred by the BOSS data. However, the method of calibrating the $\alpha_{\rm{obs}}$ estimates from the observations with the simulations is not sensitive to a constant shift in the distribution nor is it sensitive to the extremes of the magnitude distribution by construction.
     
\subsection{Calibration procedure on simulations}\label{method:sim}

\begin{figure*}
       \centering
       \includegraphics[width=14cm]{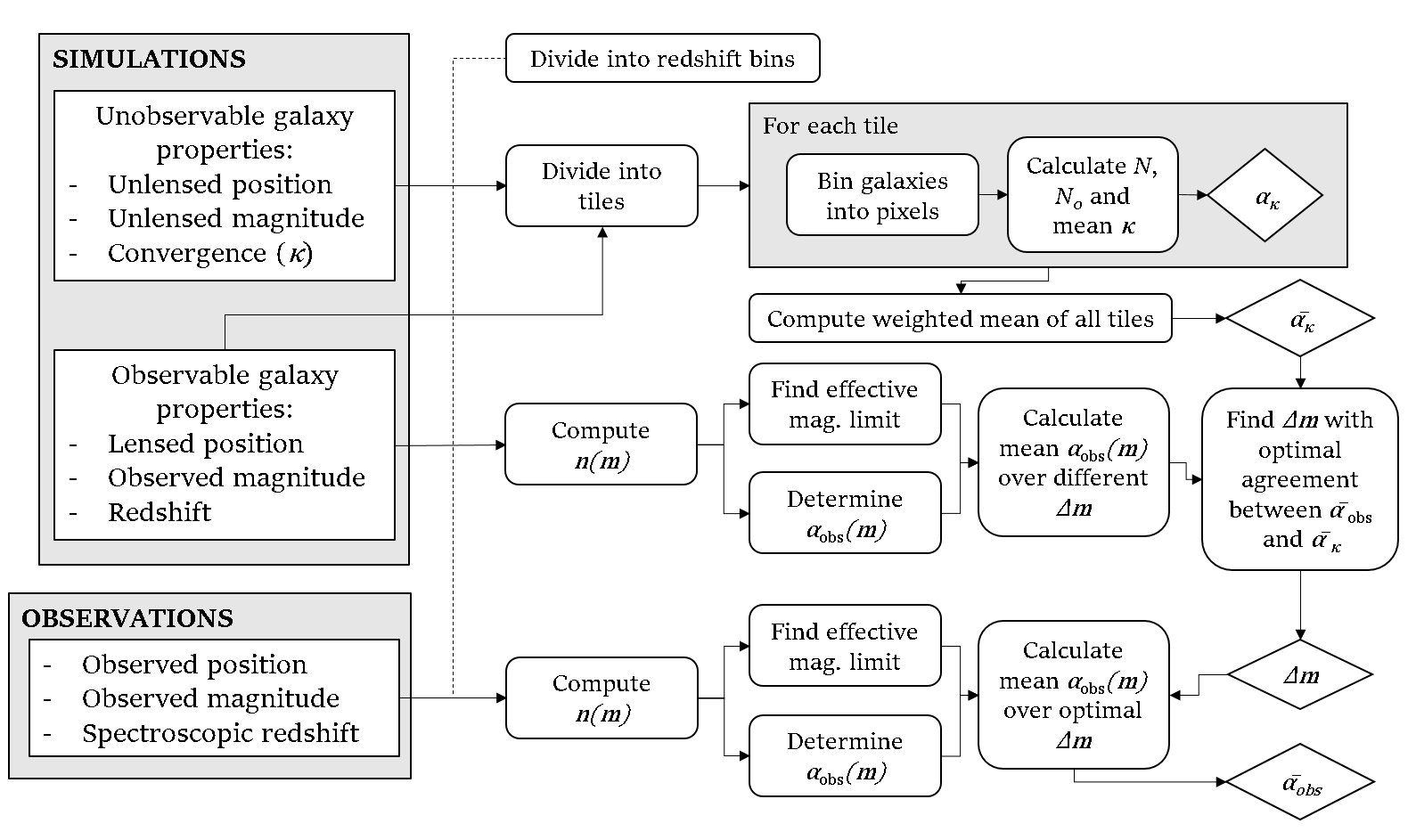}
      \caption{Flow chart outlining the method presented in this paper. $N$ stands for the count of lensed galaxies, $N_{0}$ refers to the counts of unlensed galaxies, $\kappa$ to the convergence, $\alpha_{\kappa}$ to the luminosity function slope determined from the known $\kappa$, $n(m)$ is the differential galaxy count distribution over magnitude, $m$, $\alpha_{\rm{obs}}$ is the luminosity function slope as determined from $n(m)$.}
    \label{fig:diagram}
   \end{figure*}

\begin{figure}
       \centering
       \includegraphics[width=\columnwidth]{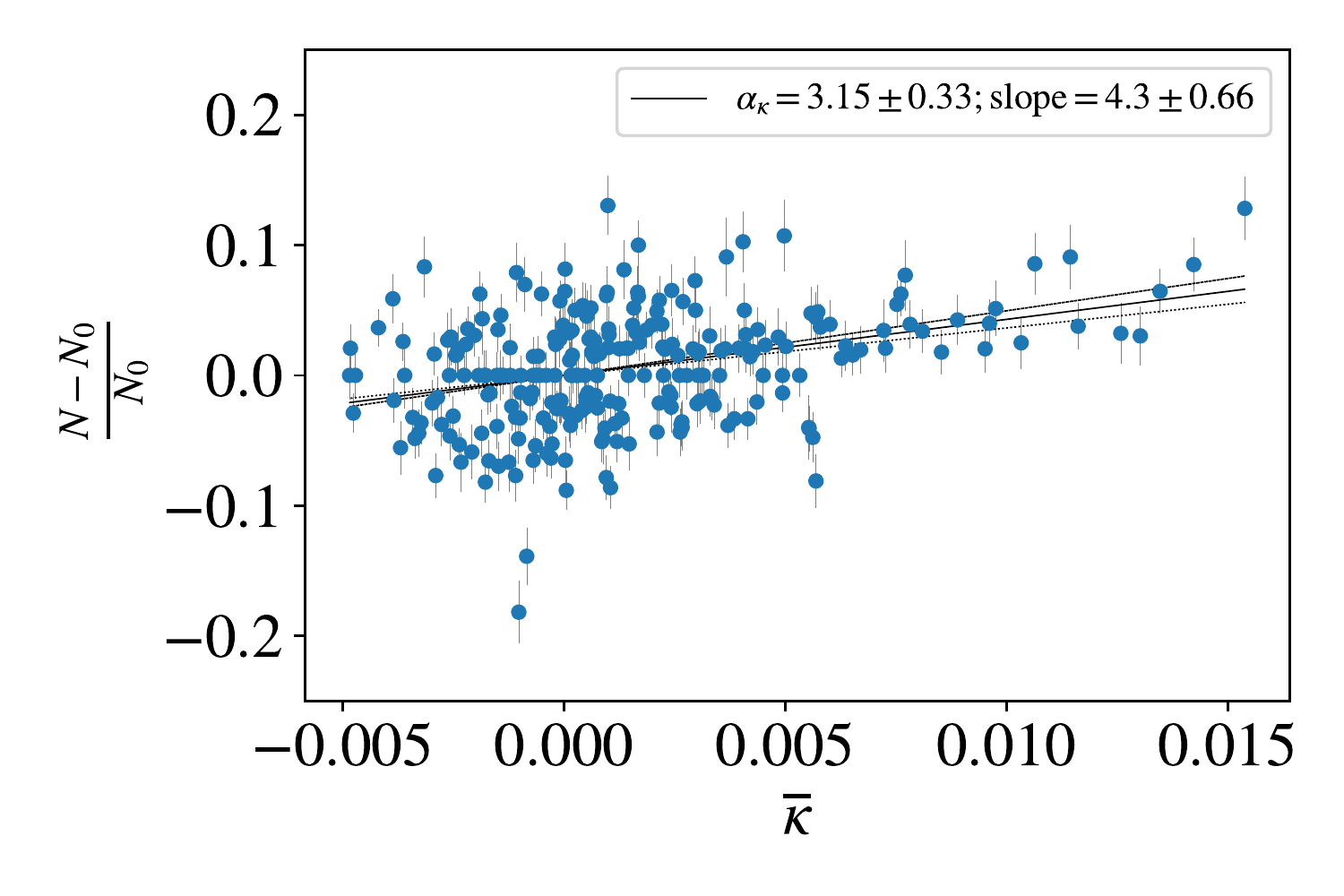}
      \caption{Plot of the relative difference in galaxy counts per pixel over the mean convergence ($\overline{\kappa}$) in each pixel for a \textsc{HealPix} pixelation with nside = 64 and in the zhigh redshift bin ($0.5 < z \leq 0.75$). The graph only shows pixels within 1 of 28 tiles. The relative difference between the lensed and unlensed galaxy counts in each pixel are shown as blue points. The black line is fitted to the blue data points with equation~(\ref{eqn:reldiff}) to give the $\alpha_{\kappa}$ value shown in the legend.}
    \label{fig:64_kappa_vs_countdiff_tile_9}
   \end{figure}

    To calibrate the $\alpha_{\rm{obs}}$ obtained from observations, we first have to determine an accurate estimate of the underlying luminosity function slope, $\alpha_{\kappa}$, in the MICE2 simulations as given by equation~(\ref{eqn:reldiff}). As outlined in figure~\ref{fig:diagram}, we first spatially bin the lensed and unlensed galaxy positions using \textsc{HealPix} at a resolution of nside = 64 \citep{gorski2005healpix}. Within each bin/pixel, we evaluate lensed and unlensed cumulative galaxy counts, $N$ and $N_{0}$ respectively, as well as the average convergence, $\kappa$. We then perform a least squares linear fit of the relative difference between lensed and unlensed galaxy counts over the convergence, $\kappa$, to estimate $\alpha_{\kappa}$ (as shown in figure~\ref{fig:64_kappa_vs_countdiff_tile_9}). This is a consequence of the linearity between these two quantities which emerges in the weak lensing limit as given by equation~(\ref{eqn:reldiff}).\\
    
    In order to obtain better estimates of the uncertainties of $\alpha$, the \textsc{HealPix} pixels are grouped into tiles (\textsc{HealPix} pixels with a resolution of nside = 4) for which we repeat the analysis independently each time. The weighted mean of these values obtained from each tile gives the final estimate for $\alpha_{\kappa}$, $\overline{\alpha_{\kappa}}$, while the standard deviation between these values is used to estimate the uncertainty as given by
    
    \begin{equation}
    	\sigma_{\rm{\alpha}}^{2} \equiv 
        	    \frac{
        	        \sum\limits_{\rm{i}=1}^{M} 
        	        \frac{
        	            (\alpha_{\rm{i}} - \overline{\alpha} )^{2}} 
        	            {\sigma_{\rm{i}}^{2}} 
        	        } 
        	        {(M-1)
        	        \sum\limits_{\rm{i}=1}^{M}
        	        \frac{1}{\sigma_{\rm{i}}^{2}}
        	        },
    	\label{eqn:unc}
    \end{equation}
    \noindent where $\alpha_{\rm{i}}$ are the $\alpha$ estimates from each tile or bin, $\sigma_{\rm{i}}$ is their associated uncertainty, $\overline{\alpha}$ is the weighted mean of the $\alpha$ estimates and $M$ is the number of tiles over which the analysis is repeated. When $\sigma_{\rm{i}} = \sigma$, equation~(\ref{eqn:unc}) reduces to the equation~for the error of the mean, i.e. $\sigma_{\alpha} \equiv \sigma_{\rm{sd}}/\sqrt{M}$ where  $\sigma_{\rm{sd}} \equiv (\sum_{\rm{i}=1}^{M} (\alpha_{\rm{i}} - \overline{\alpha} )^{2}/(M-1))^{1/2}$.\\
    
    As an alternative, one might think that it would be enough to assume that the uncertainty on the galaxy counts is given by a noise, which considers the correlation between the lensed and unlensed galaxy counts (which is shown in the errorbars of the data points in figure~\ref{fig:64_kappa_vs_countdiff_tile_9}). We find, however, that this approach leads to underestimates of the uncertainties. Sampling $\alpha_{\kappa}$ over many different areas in the sky gives a more conservative estimate of the uncertainty, while also accounting for the local fluctuations in the BOSS sample.\\
    
    A possible cause for concern when comparing the magnified and unmagnified galaxy populations can be the edge cases where, for a given bin or pixel, the unmagnified galaxy number count $N_{0} = 0$, while the magnified number counts $N = 1$ or vice versa. These cases cause di\-vergences in the relative difference and unrealistic uncertainties, since they introduce null denominators. For this reason, they are excluded in the analysis. In any case, the frequency of these occurrences is usually found to be negligible for the \textsc{HealPix} resolutions and redshift bins used in this work. Dividing the 5000 deg$^2$ MICE2 simulations into two redshift bins at a \textsc{HealPix} nside = 64, there are none of these cases. While considering 19 redshift bins at the same \textsc{HealPix} resolution, only $\mysim 0.7\%$ of the pixels have to be discarded.

\subsection{Determining magnification bias from observations}\label{method:obs}
   After having determined the luminosity function slope, $\overline{\alpha}_{\kappa}$, from the simulations as described in section~\ref{method:sim}, we estimate the optimal magnitude range, $\Delta m$, to calibrate the estimate of $\alpha_{\rm{obs}}$ from mock observations using $\alpha_{\kappa}$.\\
    
    To do this, we first choose a magnitude band, $m$, that has been used to select (at least, partially) the galaxy sample of interest. Another magnitude band will carry less information about flux magnification. Then, we determine the discrete differential galaxy count distribution, $n(m)$, over the chosen magnitude, $m$, for a given redshift range. Subsequently, we find the magnitude at which the faintest most dominant peak in $n(m)$ occurs. This value is considered to be the effective magnitude limit of the galaxy sample. From $n(m)$, we compute $\alpha_{\rm{obs}}(m)$ using equation~(\ref{eqn:alpha_obs}). Thereafter, we calculate the weighted mean of $\alpha_{\rm{obs}}(m)$, $\overline{\alpha}_{\rm{obs}}$, over all possible magnitude ranges, $\Delta m$, below the effective magnitude limit determined before.\\
    
    In order to find the optimal $\Delta m$ which will be used for the ca\-libration of $\overline{\alpha}_{\rm{obs}}$ from the actual observations, we find the value of $\overline{\alpha}_{\rm{obs}}(\Delta m)$ which is in best statistical agreement with the value of $\overline{\alpha}_{\kappa}$ determined previously for the same galaxy sample and redshift range. Therefore, the optimal $\overline{\alpha}_{\rm{obs}}(\Delta m)$ value is the one which mi\-nimises the number of standard deviations it deviates from $\overline{\alpha}_{\kappa}$, i.e. $|\overline{\alpha}_{\rm{obs}}(\Delta m) - \overline{\alpha}_{\kappa}|/\sigma_{\alpha_{\kappa}}$.\\
    
    The reason behind choosing a magnitude range, $\Delta m$, relative to the effective magnitude limit of the differential galaxy count distribution, $n(m)$, for calibration is to account for one of the simplest forms of disagreement between the observed $n(m)$ and the $n(m)$ from mock observations. This disagreement being a constant shift in the domain of $n(m)$. For instance, such a shift exists between the $n(m)$ from the BOSS and MICE2 samples which has been discussed in section~\ref{method:data} and shown in figure~\ref{fig:count_hist} already. If we were to evaluate $\overline{\alpha}_{\rm{obs}}^{\rm{MICE2}}$ and $\overline{\alpha}_{\rm{obs}}^{\rm{BOSS}}$ over the same magnitude range, while disregarding the difference between their $n(m)$ distributions, the $\overline{\alpha}_{\rm{obs}}$ estimates will be biased. This happens because we would be probing regimes of $n(m)$ from the observed galaxy sample beyond or far below its magnitude limit when calculating $\overline{\alpha}_{\rm{obs}}$. Other higher-order biases in the $n(m)$ from mock observations may exist which would require more complex parametrisations of the calibration procedure. Nevertheless, in such cases, it might be more efficient and physically motivated to adjust the models used to produce the mock galaxy samples such that the agreement in $n(m)$ improves up to a point where it can be mostly parametrised by a constant shift in the magnitude.\\
    
    In any case, once the optimal $\Delta m$ to reconcile $\overline{\alpha}_{\rm{obs}}$ and  $\overline{\alpha}_{\kappa}$ from the mocks has been determined, it may be used to calibrate $\overline{\alpha}_{\rm{obs}}$ from the observations. As summarised in the lower third of figure~\ref{fig:diagram}, we first compute $n(m)$ for the given redshift range. We again find the faintest most dominant peak in $n(m)$ and set it as the effective magnitude limit and evaluate $\alpha_{\rm{obs}}(m)$ from $n(m)$. Lastly, we calculate the weighted mean of $\alpha_{\rm{obs}}(m)$ over the optimal magnitude range below the effective magnitude limit, $\Delta m$, determined before from the simulations over the same redshift range. Thus, we produce the final $\overline{\alpha}_{\rm{obs}}$ estimate for that sample.

\section{Applications to BOSS lenses} \label{non-flux-lim}
    We proceed to apply the method described in sections~\ref{method:sim} and \ref{method:obs} to the BOSS lens galaxy sample introduced in section~\ref{method:boss_data}. The magnitude bands selected for this are cmodel magnitudes, since they are better indicators of the overall flux emitted by a galaxy. The specific magnitude band chosen is based on which band was used to select the dominant population within a sample. In other words, when working with LOWZ-dominated galaxy samples ($z < 0.36$), we use the $r$-band and when working with CMASS-dominated samples ($z > 0.36$), we use the $i$-band \citep{eisenstein2011sdss}. To allow for accurate forecasting of the KiDS-1000+BOSS analysis \citep{heymans2020kids}, we choose the same convention for the redshift bins: $0.2 < z \leq 0.5$ and $0.5 < z \leq 0.75$. Consequently, both bins are dominated by CMASS galaxies, so we opt to use $i$-band magnitudes for the analysis of both samples.\\

    As demonstrated in appendix~\ref{flux-lim}, for the flux-limited case, we can accurately and robustly estimate the magnitude of the magnification bias by determining the effective luminosity function slope $\alpha$ through the weighted mean of $\alpha_{\rm{obs}}$ near the magnitude limit. In this section, we discuss whether the same can be said when applying a complex sample selection function which does not have a clear flux/magnitude limit such as in the case of the BOSS survey.\\
    
    Firstly, we directly estimate $\alpha_{\kappa}$ from the MICE2 simulations following the approach outlined in section~\ref{method:sim}. An example of this is shown in figure~\ref{fig:64_kappa_vs_countdiff_tile_9}, where we see the $\alpha_{\kappa}$ estimate within a single $\mysim200$ deg$^2$ tile containing 256 pixels within the zhigh bin. This procedure is repeated for each tile and redshift bin. Then, we find the weighted mean between the $\alpha_{\kappa}$ from each tile to determine the $\overline{\alpha}_{\kappa}$ for each redshift bin and its uncertainty given by equation~(\ref{eqn:unc}). This gives $\overline{\alpha}^{\rm{zlow}}_{\kappa} = 2.43 \pm 0.09$ and $\overline{\alpha}^{\rm{zhigh}}_{\kappa} = 3.26 \pm 0.07$.\\
    
\begin{figure}
        \centering
        \includegraphics[width=\columnwidth]{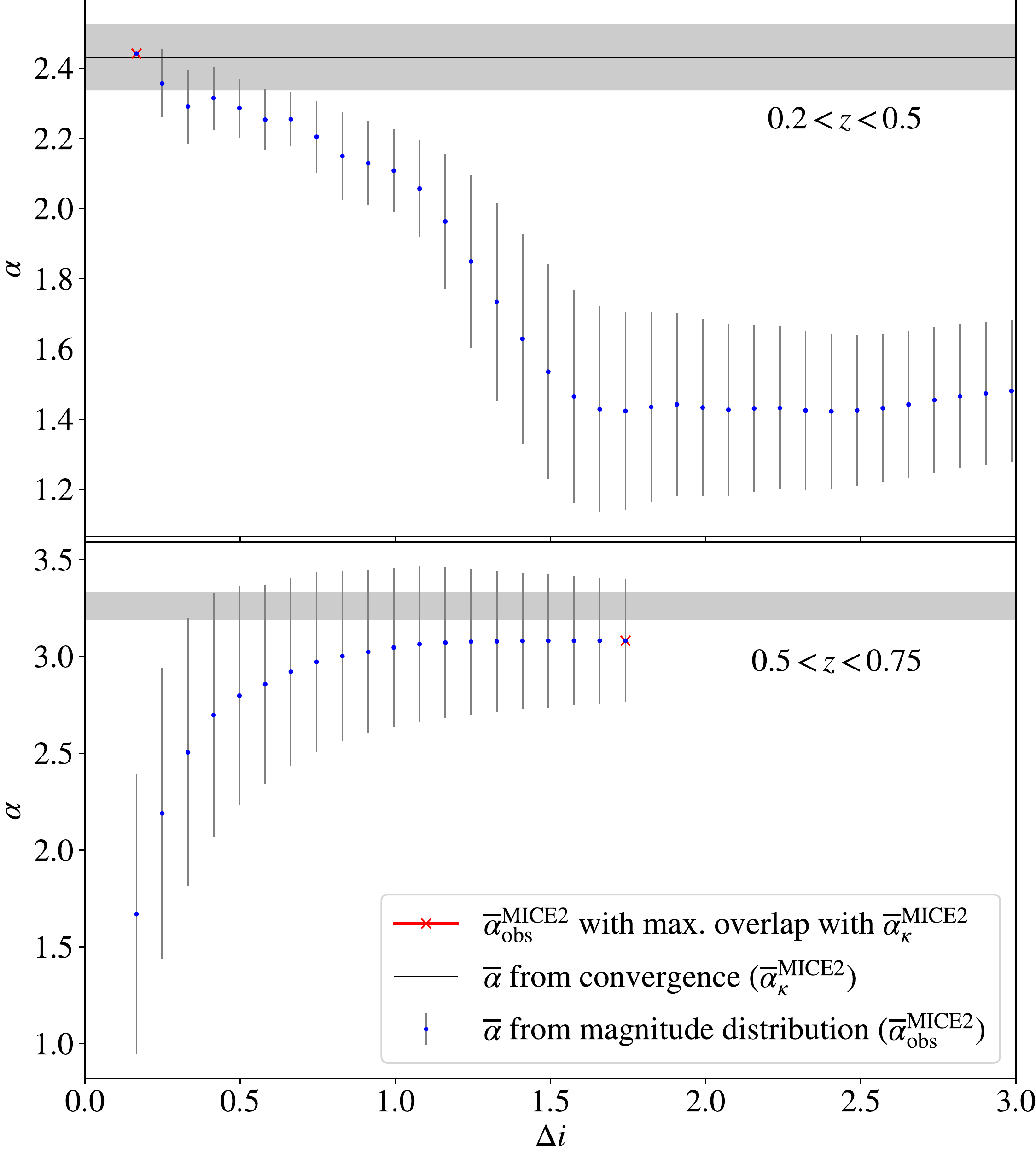}
        \caption{$\overline{\alpha}_{\rm{obs}}$ estimates from MICE2 simulations with the BOSS selection function over different $i$-band magnitude ranges below the turn-off magnitude, $\Delta i$, considered to calculate the weighted average. Two redshift ranges are considered: zlow with $0.2 < z \leq 0.5$ (top) and zhigh with $0.5 < z \leq 0.75$ (bottom). The red cross marks the $\alpha$ estimate which overlaps the most with the $\overline{\alpha}_{\kappa}$ truth from the simulations (black vertical line).}
        \label{fig:alpha_varying}
    \end{figure}

    Next, applying the procedure discussed in section~\ref{method:obs} and using the differential galaxy count distributions for each redshift bin shown in figure~\ref{fig:count_hist}, we can estimate $\overline{\alpha}_{\rm{obs}}$; once for the simulated BOSS-MICE2 observations, and once for the actual BOSS observations. In figure~\ref{fig:alpha_varying}, for zlow, we find that the estimate is optimal near the faint end of the count distribution, which is expected, since the assumed flux power law should be most accurate in the faint limit. However, this does not appear to be the case for the high redshift sample, zhigh. For this range, the estimate is optimal when considering the whole magnitude range up to the turn-off magnitude. This might be due to incompleteness in the sample and/or the complex selection, which flattens the observed number counts \citep{hildebrandt2016observational}.\\

\begin{figure}
       \centering
       \includegraphics[width=\columnwidth]{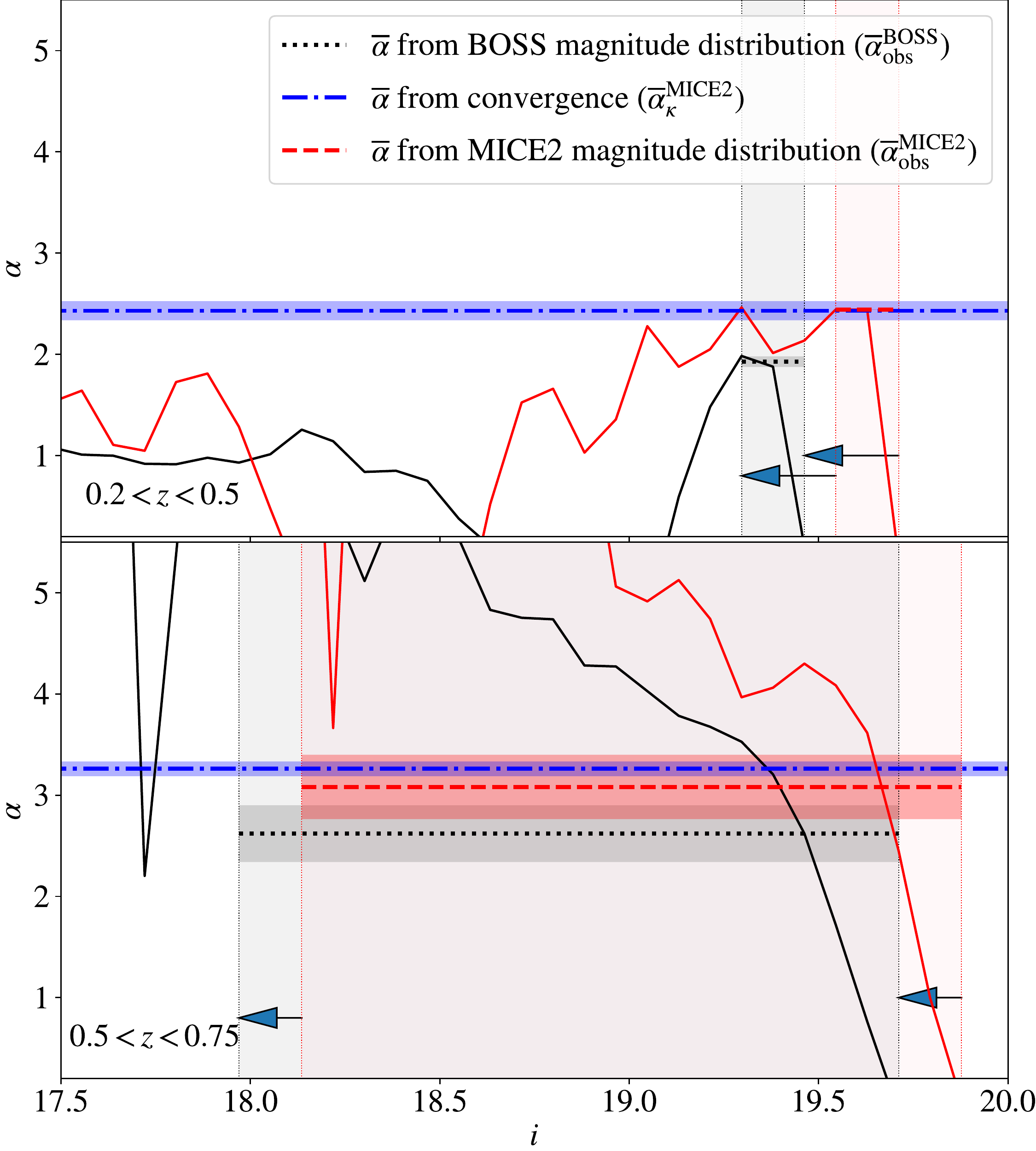}
          \caption{The slope of the BOSS luminosity function, $\alpha$, as a function of the $i$-band magnitude (i) for two redshift bins:  $0.2 < z \leq 0.5$ (top) and $0.5 < z \leq 0.75$ (bottom). The red line shows $\alpha_{\rm{obs}} (i)$ as given by equation~(\ref{eqn:alpha_obs}) calculated from the MICE2 mocks, while the black line shown $\alpha_{\rm{obs}} (i)$ as determined from the BOSS DR12 photometric data. The grey vertical lines mark the upper and the lower bounds of the magnitude range used to find $\overline{\alpha}_{\rm{obs}}^{\rm{BOSS}}$, while the red vertical lines mark the upper and lower bounds of the highlighted magnitude range  used to determine $\overline{\alpha}_{\rm{obs}}^{\rm{MICE2}}$. The arrows indicate the constant magnitude shift applied to reconcile the differential galaxy count distribution, $n(m)$, from observations with the $n(m)$ from mocks. The dotted black horizontal line marks the $\overline{\alpha}_{\rm{obs}}$ estimate from BOSS galaxies, the
          dashed red horizontal line marks the $\overline{\alpha}_{\rm{obs}}$ estimate from MICE2 mock galaxies and the blue dot-dashed horizontal line marks the effective $\overline{\alpha}^{\rm{MICE2}}_{\kappa}$ determined from the weak lensing convergence with equation~(\ref{eqn:reldiff}) and used to calibrate $\overline{\alpha}_{\rm{obs}}^{\rm{MICE2}}$.}
         \label{fig:alpha_hist1}
   \end{figure}

    Taking the magnitude range from the optimal $\overline{\alpha}^{\rm{MICE2}}_{\rm{obs}}$ estimate to calibrate $\overline{\alpha}^{\rm{BOSS}}_{\rm{obs}}$ gives the estimates shown in figure~\ref{fig:alpha_hist1}. For the MICE2 mocks, we find that $\overline{\alpha}^{\rm{zlow}}_{\kappa} = 2.43 \pm 0.09$, while $\overline{\alpha}_{\rm{obs}}^{\rm{zlow}} = 2.442 \pm 0.002$. In addition, $\overline{\alpha}^{\rm{zhigh}}_{\kappa} = 3.26 \pm 0.07$, while $\overline{\alpha}_{\rm{obs}}^{\rm{zhigh}} = 3.08 \pm 0.32$ which indicates that the $\alpha$ estimates obtained from observations using equation~(\ref{eqn:alpha_obs}) are a good indicator of the scale of the magnification bias even when there is a complex sample selection function when they are properly calibrated. For this reason, we may consider the $\overline{\alpha}_{\rm{obs}}$ estimates given in table~\ref{table:BOSS-results} from the actual BOSS observations as unbiased indicators of the scale of the magnification bias. Note that the value for zlow, slightly deviates from the value of $\overline{\alpha}_{\rm{obs}}^{\rm{BOSS}} = 1.80 \pm 0.15$ quoted in \cite{joachimi2020kids}, since there have been minor adjustments in the way peaks in $n(m)$ are detected. This leads to a 16\% change in the amplitude of the mag. bias contribution, which has no effect on the KiDS-1000 analysis as the GGL contributions are marginal.\\
    
    When comparing the $\alpha_{\rm{obs}}(i)$ curves for each bin in figure~\ref{fig:alpha_hist1}, one might notice that the turn-off near the effective magnitude limit is not as steep for zhigh as for zlow. This is due to the complex BOSS selection function which deviates particularly strongly from a simple flux limit at high redshifts. Here is where the semi-empirical calibration of the magnitude range considered in order to determine the effective luminosity function slope $\overline{\alpha}_{\rm{obs}}$ is especially relevant. As shown in figure~\ref{fig:alpha_varying}, we find that for zhigh we get a more accurate $\alpha$ estimate when considering the entire magnitude range $\Delta i$ available below the effective magnitude limit which is in stark contrast with the results found for a flux-limited sample (see figure~\ref{fig:flux-lim-alpha-varying}). The opposite is the case for zlow. As shown in figure~\ref{fig:alpha_hist1}, the double peak in the zlow bin combined with a clearer 'flux limit' near the peak magnitude means that the power law model for the luminosity function holds best within a small magnitude range near the peak. In other words, the $\Delta i$ intervals which provide the best agreement between $\overline{\alpha}_{\rm{obs}}$ and $\overline{\alpha}_{\kappa}$ are also the ma\-gnitude intervals over which $n(m)$ resembles a power law the most. Therefore, our method actively avoids basing its estimates on a magnitude domain where the power law approximation in equation~\ref{eqn:n0} does not hold.\\ 
    
    We note that in figure~\ref{fig:count_hist} the simulated and the observed diffe\-rential count distributions do not quite match. The $n(i)$ from MICE2 mock observations is shifted by a $\Delta i \approx 0.2$ to the faint end with respect to the BOSS $n(i)$. This might be due to some limitations in the galaxy model of the MICE2 simulations. The fact that the $n(m)$ from the mocks and observations do not match perfectly seems to be driving the discrepancy between $\overline{\alpha}_{\rm{obs}}^{\rm{MICE2}}$ and $\overline{\alpha}_{\rm{obs}}^{\rm{BOSS}}$ shown in figure~\ref{fig:alpha_hist1}. However, since our calibration is based on a magnitude range of a fixed width relative to the effective magnitude limit for each sample, the estimates are not sensitive to this apparent shift in the domain of $n(i)$. The only thing which can bias our estimates are any disagreements in higher-order derivatives of $n(m)$ near the effective magnitude limit between observations and simulations. However, the uncertainties of $\overline{\alpha}_{\rm{obs}}$ from equation~(\ref{eqn:unc}) are defined such that they consider the variations of $\overline{\alpha}_{\rm{obs}}$ within the calibration magnitude range.\\

\begin{figure*}
       \centering
       \includegraphics[width=14.5cm]{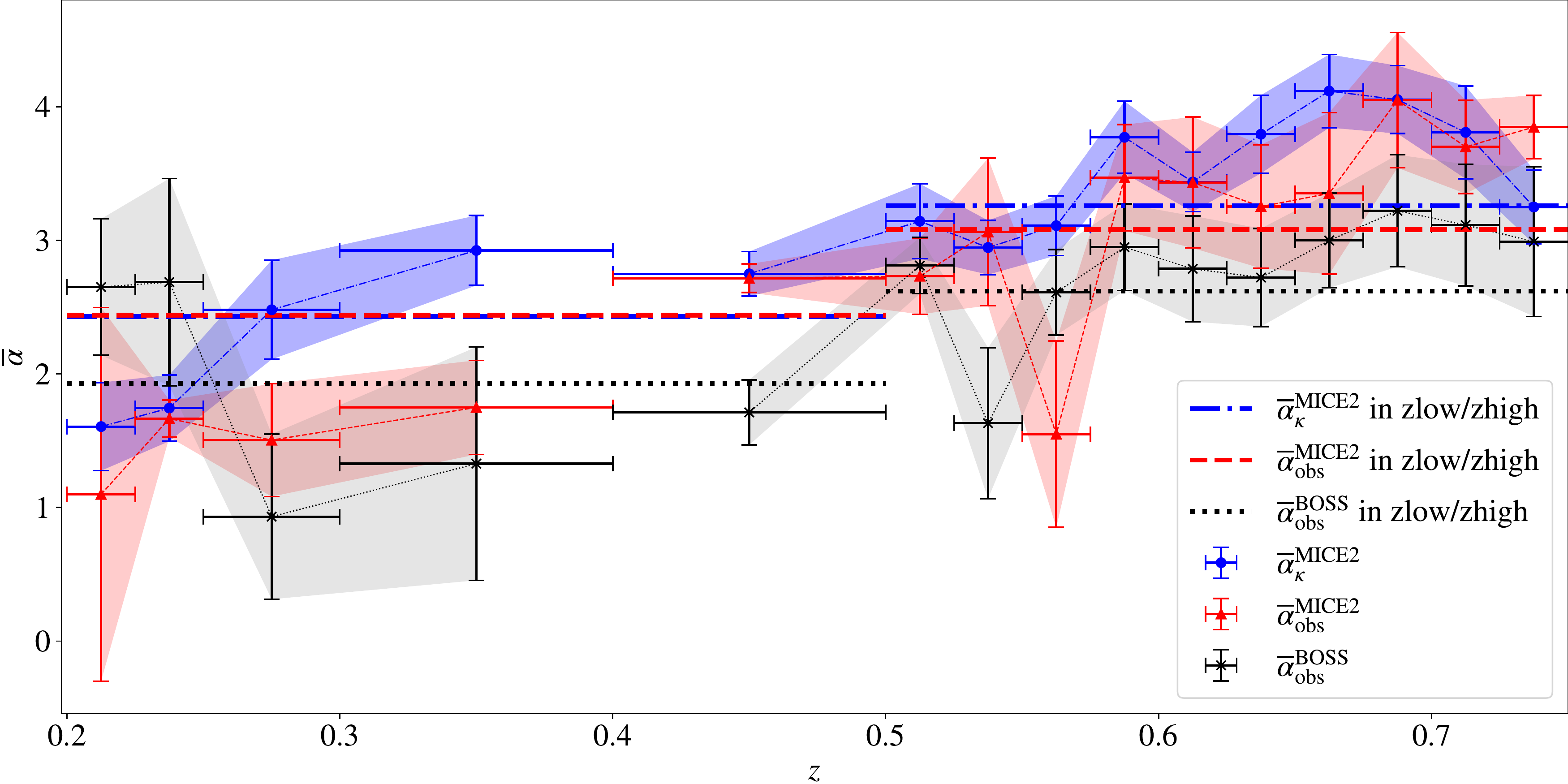}
      \caption{Plot of different $\alpha$ estimates for 16 different redshift ($z$) bins within $0.2 < z \leq 0.75$. The black crosses mark the $\overline{\alpha}_{\rm{obs}}^{\rm{BOSS}}$ estimates from observations within each bin, the red triangles mark $\overline{\alpha}_{\rm{obs}}^{\rm{MICE2}}$ estimates from mock observations and the blue circles mark the true effective $\overline{\alpha}^{\rm{MICE2}}_{\kappa}$ determined from the weak lensing convergence with equation~(\ref{eqn:reldiff}) and used to calibrate $\overline{\alpha}_{\rm{obs}}^{\rm{MICE2}}$.The values of $\overline{\alpha}_{\rm{obs}}^{\rm{MICE2}}$ and $\overline{\alpha}_{\rm{obs}}^{\rm{BOSS}}$ for the bins with $z < 0.4$ have been derived from the differential galaxy count distribution with respect to the $r$-band magnitude, $n(r)$, while the values for the bins with $z > 0.4$ have been derived from $n(i)$. The horizontal lines show the $\alpha$ estimates from simulations obtained for the zlow bin ($0.2 < z \leq 0.5$) and the zhigh bin ($0.5 < z \leq 0.75$).}
    \label{fig:many_z_bins}
   \end{figure*}

    In addition, to see how $\alpha$ evolves over redshift within zlow and zhigh, we repeat this analysis of the BOSS sample again for a different choice of redshift bins producing the $\alpha$ estimates shown in figure~\ref{fig:many_z_bins}. Here, the edges of the 15 redshift bins are given by $\{0.2, 0.225, 0.25, 0.3, 0.4, 0.5, 0.525,0.55, 0.575, 0.6,$ $0.625, 0.65, 0.675, 0.7, 0.725, 0.75\}$. Since the redshift bins between $z = 0.2$ and $z = 0.4$ are dominated by LOWZ galaxies, we choose a bin width of 0.1 instead of 0.025 between $z = 0.3$ and $z = 0.5$. This is done to mitigate the sharp gradient changes in $n(m)$ in the BOSS sample at redshifts near $z = 0.36$, i.e. at the boundary between the LOWZ and CMASS samples as shown in figure~\ref{fig:counts}.\\

    Figure~\ref{fig:many_z_bins} shows how the effective luminosity function slope $\overline{\alpha}_{\kappa}$ in the MICE2 sample varies smoothly. Nonetheless, $\overline{\alpha}_{\rm{obs}}$ for MICE2 and for BOSS varies more strongly with redshift, due to their sensitivity of small variations in $n(m)$. Also, $\overline{\alpha}_{\kappa}^{\rm{MICE2}}$ is consistent with $\overline{\alpha}_{\rm{obs}}^{\rm{MICE2}}$ over most of the redshift range. However, for a few redshift bins, $\overline{\alpha}_{\rm{obs}}^{\rm{MICE2}}$ is in a $\mysim1 \sigma$ to $\mysim2 \sigma$ tension with $\overline{\alpha}_{\kappa}$ despite being calibrated to optimally overlap. Taking $\overline{\alpha}_{\kappa}$ as the underlying truth, we consider $\overline{\alpha}_{\rm{obs}}^{\rm{MICE2}}$ and  $\overline{\alpha}_{\rm{obs}}^{\rm{BOSS}}$ to be biased in these cases. This seems to be driven by small discrepancies between the faint-end of $n(m)$ from MICE2 and the faint-end of $n(m)$ from BOSS. These are then exacerbated, since a small change in the sample size can lead to radical changes in the gradient of the magnitude distribution $n(m)$ of these galaxies, causing substantial biases in the $\overline{\alpha}_{\rm{obs}}$ estimates, as discussed in \cite{hildebrandt2016observational}. Nonetheless, these discrepancies become insignificant as we increase the sample size by widening the redshift bin width to the one used in the main analysis (i.e. $0.2 < z \leq 0.5$ and $0.5 < z \leq 0.75$). We also note that the $\overline{\alpha}_{\rm{obs}}^{\rm{BOSS}}$ estimates for the $0.2 < z \leq 0.225$ and $0.225 < z \leq 0.25$ bins may be biased. This is the case, since the profile of $n(m)$ as obtained from the MICE2 simulations deviates from the $n(m)$ observed in BOSS more strongly than over the remaining redshift range. Hence, the calibration range determined through our method does not necessarily apply anymore (as already mentioned in section~\ref{method:obs}) and the estimates may be inaccurate. To avoid this, we highlight the necessity for accurate cosmological simulations over the whole redshift domain.

    \begin{table}
        \caption{Table showing the effective luminosity function slopes derived from figure~\ref{fig:alpha_hist1} for each redshift bin of the BOSS galaxy sample.}
        
        \begin{tabular}{lP{2.0cm}P{4.5cm}}
        
        \hline
        Bin & Redshift range & Luminosity function slope ($\overline{\alpha}_{\rm{obs}}^{\rm{BOSS}}$) \\
        \hline
        zlow           & $0.2 < z \leq 0.5$         & $1.93 \pm 0.05$              \\
        zhigh           & $0.5 < z \leq 0.75$       & $2.62 \pm 0.28$              \\
        \hline
        \end{tabular}
        
        \label{table:BOSS-results}
    \end{table}

\section{Magnification bias in weak lensing measurements}\label{forecasts}

\begin{figure}
       \centering
       \includegraphics[width=7.2cm]{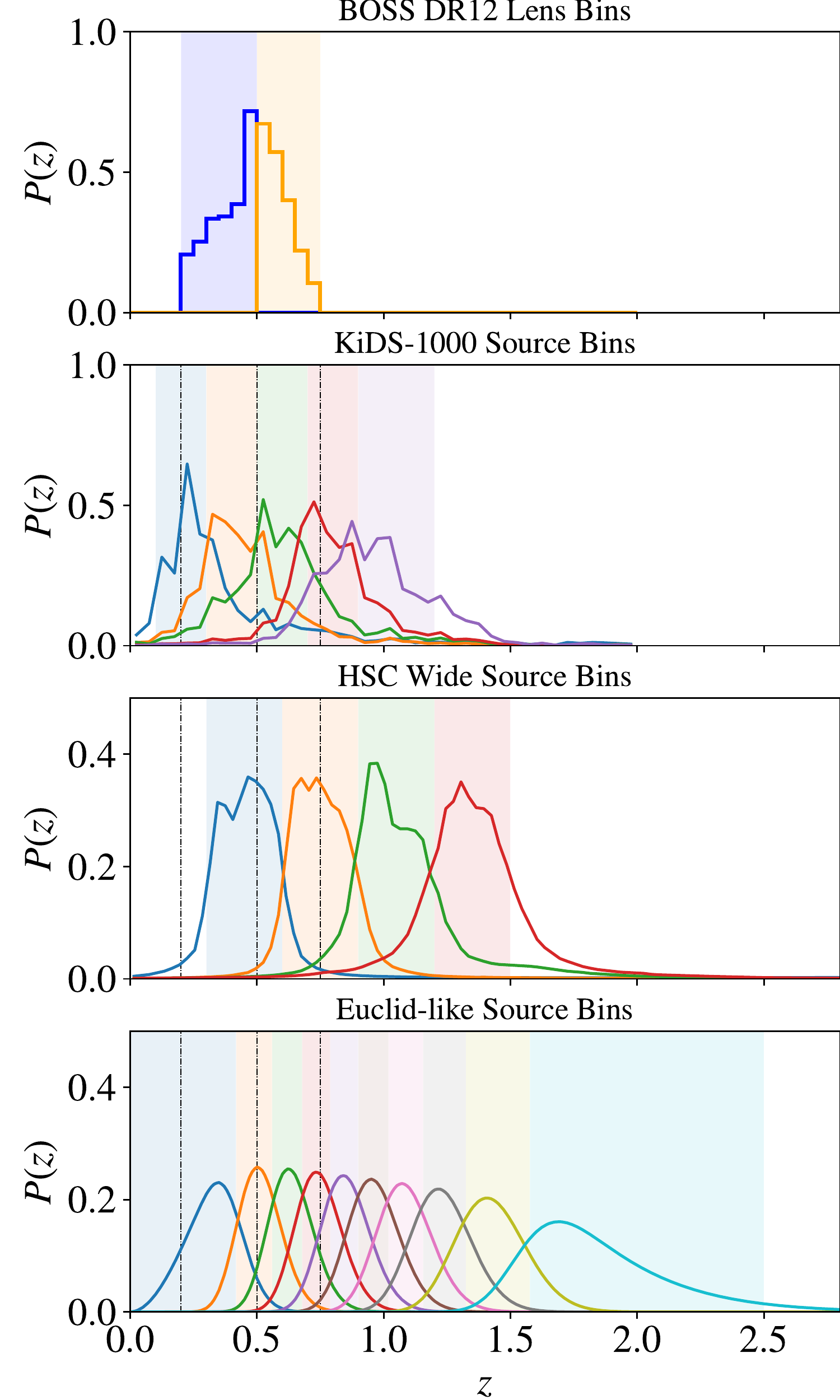}
      \caption{Redshift distributions $P(z)$ for the lens and source galaxy samples used in the forecasts for the galaxy-galaxy lensing signal in a KiDS-1000+BOSS, HSC Wide+BOSS and \textit{Euclid}-like+DESI-like analysis. The pro\-perties of these redshift distributions are given in table~\ref{table:forecasts}.The black vertical dot-dashed lines show the limits of the BOSS lens bins for comparison with the source bins.}
    \label{fig:all_nofz}
   \end{figure}

    Having produced estimates for the effective luminosity function slope ($\overline{\alpha}_{\rm{obs}}$) for the BOSS DR12 galaxy sample, we now proceed to make forecasts of the importance of magnification bias in the GGL signals. The forecasts are produced from cross correlating source galaxies from weak lensing surveys with the BOSS lens samples considered in section~\ref{non-flux-lim}. First, we produce forecasts for the GGL signals for a KiDS-1000+BOSS DR12 analysis as described in \cite{joachimi2020kids}. Secondly, we produce similar forecasts for a GGL analysis of HSC Wide+BOSS DR12 similar to \cite{speagle2019galaxy}, while using the source bins described in \cite{hikage2019cosmology}. Lastly, we produce GGL forecasts for a potential \textit{Euclid}-like+DESI-like analysis using the galaxy sample properties defined in the \textit{Euclid} collaboration forecast choices \citep{blanchard2019euclid}, The properties of all of the aforementioned galaxy samples are given in table~\ref{table:forecasts} and their redshift distributions, $P(z)$, are given in figure~\ref{fig:all_nofz}.\\
    
    Throughout the forecasts, we assume a Planck 2018 TT,TE,EE+lowE flat $\Lambda$CDM cosmology \citep{aghanim2020planck} with $\omega_{\rm{b}} = 0.02236$, $\omega_{\rm{c}} = 0.1202$, $h = 0.6727$, $n_{\rm{s}} = 0.9649$, $\rm{ln}(10^{10}A_{\rm{s}}) = 3.045$, $\Omega_{\kappa} = 0$, $w = -1$, and $\sum m_{\nu} = 0.06$ eV c$^{-2}$. To model the cross-power spectrum between galaxy and matter distribution ($P_{\rm{gm}}$, e.g. section~\ref{back:signal}), we split the power spectrum into linear and a non-linear part as outlined in \cite{joachimi2020kids} based on \cite{sanchez2017clustering} and set $b_{1} = [2.1, 2.3]$, $b_{2} = [0.2, 0.5]$, and $\gamma_{3} = [0.9, 0.1]$ where the first value of each vector corresponds to the first lens bin (zlow) and the second values to the second lens bin (zhigh). These values follow the rounded best-ﬁt values from the cosmic shear and GGL analysis of KV450+BOSS \citep{troster2019cosmology}. We use the halo and intrinsic alignment models described in section~\ref{background} and set $A_{\rm{bary}} = 3.13$ (upper limit of the KiDS-1000 prior) and $A_{\rm{IA}} = 0.8$ (best estimate from \citealt{troster2019cosmology}).

    \begin{table}
        \caption{Properties of the galaxy samples used to produce the galaxy-galaxy lensing forecasts.}
        \begin{tabular}{lP{2.8cm}lllll}
        \hline
        Bin & $z$ range & $\overline{z}$ & $z_{\rm{med}}$ & $n_{\rm{gal}}$ & $\sigma_{\epsilon, i}$ \\
        \hline
        zlow           & $0.2 < z_{\rm{spec}} \leq 0.5$         & 0.38 & 0.37 & 0.014 & -  \\
        zhigh           & $0.5 < z_{\rm{spec}} \leq 0.75$       & 0.60 & 0.55 & 0.016 & -   \\
        \hline
        KiDS1           & $0.1 < z_{\rm{phot}} \leq 0.3$       & 0.26 & 0.21 & 0.62  & 0.27   \\
        KiDS2           & $0.3 < z_{\rm{phot}} \leq 0.5$       & 0.40 & 0.36 & 1.18  & 0.26   \\
        KiDS3           & $0.5 < z_{\rm{phot}} \leq 0.7$       & 0.56 & 0.54 & 1.85  & 0.27   \\
        KiDS4           & $0.7 < z_{\rm{phot}} \leq 0.9$       & 0.79 & 0.75 & 1.26  & 0.25   \\
        KiDS5           & $0.9 < z_{\rm{phot}} \leq 1.2$       & 0.98 & 0.93 & 1.31  & 0.27   \\
        \hline
        HSC1           & $0.3 < z_{\rm{phot}} \leq 0.6$       & 0.61 & 0.45 & 5.5  & 0.28   \\
        HSC2           & $0.6 < z_{\rm{phot}} \leq 0.9$       & 0.78 & 0.72 & 5.5  & 0.28   \\
        HSC3           & $0.9 < z_{\rm{phot}} \leq 1.2$       & 1.09 & 1.01 & 4.2  & 0.29   \\
        HSC4           & $1.2 < z_{\rm{phot}} \leq 1.5$       & 1.37 & 1.30 & 2.4  & 0.29   \\
        \hline
         Euclid1          & $0.001 < z_{\rm{phot}} \leq 0.418$       & 0.33 & 0.21 & 3.0  & 0.21    \\
         Euclid2          & $0.418 < z_{\rm{phot}} \leq 0.560$       & 0.51 & 0.49 & 3.0  & 0.21    \\
         Euclid3          & $0.560 < z_{\rm{phot}} \leq 0.678$       & 0.63 & 0.62 & 3.0  & 0.21    \\
         Euclid4          & $0.678 < z_{\rm{phot}} \leq 0.789$       & 0.75 & 0.73 & 3.0  & 0.21    \\
         Euclid5          & $0.789 < z_{\rm{phot}} \leq 0.900$       & 0.85 & 0.84 & 3.0  & 0.21    \\
         Euclid6          & $0.900 < z_{\rm{phot}} \leq 1.019$       & 0.96 & 0.96 & 3.0  & 0.21    \\
         Euclid7          & $1.019 < z_{\rm{phot}} \leq 1.155$       & 1.09 & 1.09 & 3.0  & 0.21    \\
         Euclid8          & $1.155 < z_{\rm{phot}} \leq 1.324$       & 1.23 & 1.24 & 3.0  & 0.21    \\
         Euclid9          & $1.324 < z_{\rm{phot}} \leq 1.576$       & 1.42 & 1.45 & 3.0  & 0.21    \\
         Euclid10         & $1.576 < z_{\rm{phot}} \leq 2.500$       & 1.85 & 2.04 & 3.0  & 0.21    \\
         \hline
        \end{tabular}
       
        \small
            \textbf{Notes.} $\overline{z}$ stands for the mean redshift in each tomographic bin, $z_{\rm{med}}$ for the median redshift, $n_{\rm{gal}}$ for the galaxy number density in arcmin$^{-2}$ following the definition from \cite{heymans2012cfhtlens} and $\sigma_{\epsilon, i}$ for the dispersion per ellipticity component. zlow and zhigh are the lens bins based on the BOSS DR12 galaxy clustering data. The KiDS source bins have been defined in accordance with the methodology for the KiDS-1000 GGL analysis as given in  \cite{joachimi2020kids} and \cite{heymans2020kids} based on the redshift calibration described in \cite{hildebrandt2020kids} and \cite{wright2020photometric}. The properties of the HSC source bins are based on the information provided in table~1 of the HSC Y1 cosmic shear analysis \citep{hikage2019cosmology} and the source $P(z)$ distributions are based on the DEmP photometric redshifts. The tomographic bins for the Euclid forecasts are in accordance with the \textit{Euclid} collaboration forecast choices \citep{blanchard2019euclid}. The \textit{Euclid} $P(z)$ distributions are determined using the fitting formula from \cite{joachimi2010simultaneous} assuming equi-populated binning with an overall median redshift of 0.8.
        \label{table:forecasts}
    \end{table}
    
\subsection{KiDS-1000 + BOSS forecasts} \label{forecasts:kids}
\begin{figure*}
       \centering
       \includegraphics[width=16cm]{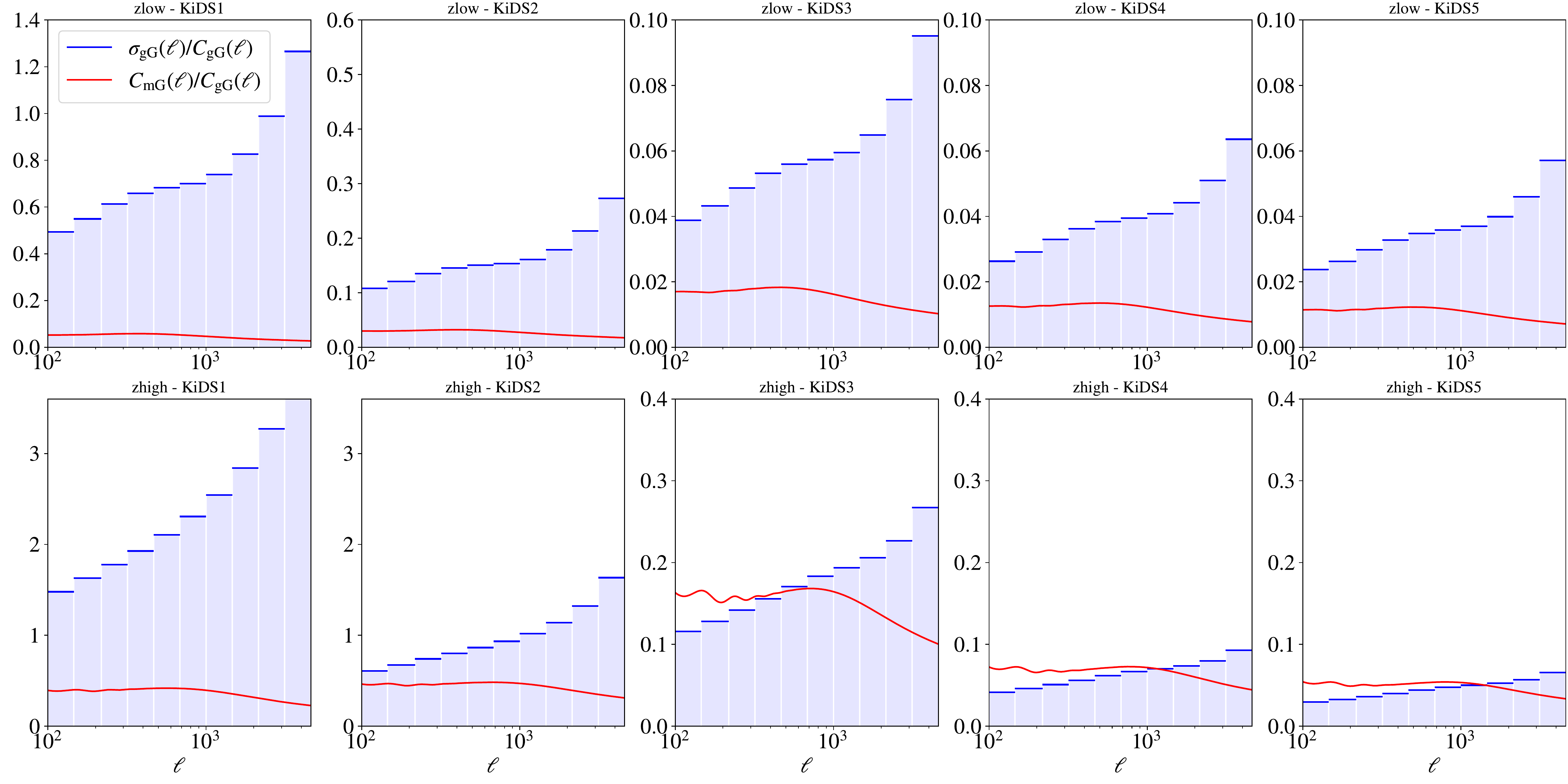}
      \caption{Magniﬁcation bias contribution $C_{\rm{mG}}(\ell)$ relative to the galaxy-galaxy lensing signal $C_{\rm{gG}}(\ell)$ over the angular scale $\ell$ (in red) for the crosscorrelations between the BOSS DR12 lens bins and the KiDS-1000 source bins assuming $\overline{\alpha}_{\rm{obs}}^{\rm{zlow}} = 1.93$ and $\overline{\alpha}_{\rm{obs}}^{\rm{zhigh}} = 2.62$. In blue, we show the expected relative uncertainty from shot and shape noise in the GGL signal, $\sigma_{\rm{gG}}(\ell)/C_{\rm{gG}}(\ell)$, within each $\ell$ bin (6 logarithmically spaced $\ell$ bins per dex). The uncertainties are calculated for a KiDS footprint with an area of 1350 deg$^2$. The properties of the galaxy samples are given in table~\ref{table:forecasts}.}
    \label{fig:kids_ggl_forecasts}
   \end{figure*}

    Following the approach outlined in section~\ref{back:signal}, we propagate the $\overline{\alpha}_{\rm{obs}}$ measurements for zlow and zhigh shown in table~\ref{table:BOSS-results} into angular power spectrum prediction for the galaxy-galaxy lensing signal. We then determine the ratio between the angular power spectrum correlating gravitational shear with the lensing-induced magniﬁcation bias in the lens sample, $C^{\rm{(i j)}}_{\rm{mG}}(\ell)$, and the angular power spectrum correlating the lens galaxy distribution and the source gravitational shear, $C^{\rm{(i j)}}_{\rm{gG}}(\ell)$, as shown in figure~\ref{fig:kids_ggl_forecasts}.\\
    
    In order to put these contributions into perspective, we also estimate the statistical uncertainty in the GGL signal assuming shot and shape noise only \citep[see for example][]{joachimi2010simultaneous}. We calculate this for 6 logarithmically spaced $\ell$ bins per dex, while assuming the footprint area of the full KiDS survey, $A = 1350$ deg$^2$. In figure~\ref{fig:kids_ggl_forecasts}, we then compare the relative magnification-shear signal to the relative GGL uncertainty for each $\ell$ bin. The magniﬁcation-shear correlation found between these bins constitutes a few-per cent contribution to the galaxy-galaxy lensing signal correlated with the zlow bin. To compare that to the shape and shot noise, $\sigma_{\rm{gG}}$, we define the cumulative signal-to-noise ratio, SNR, within a range of angular scale, $\ell_{\rm{min}} < \ell < \ell_{\rm{max}}$, as follows
    
    \begin{multline}
        \rm{SNR}(\ell_{\rm{min}} < \ell < \ell_{\rm{max}}) = \Bigg( \frac{1}{\mathit{K}} \sum^{\mathit{K}}_{\rm{i}} \rm{SNR}_{\rm{i}}^2 \Bigg)^{1/2} \\
        =  \Bigg( \frac{1}{K} \sum^{K}_{\rm{i}} \frac{C_{\rm{mG}}^2(\ell_{\rm{min, i}} < \ell < \ell_{\rm{max, i}})}{\sigma_{\rm{gG}}^2(\ell_{\rm{min, i}} < \ell < \ell_{\rm{max, i}})} \Bigg)^{1/2},
    \end{multline}
    \noindent  where $K$ is the number of $\ell$ bins, $i$ labels each $\ell$ bin, and $\ell_{\rm{min, i}}$ and $\ell_{\rm{max, i}}$ mark the lower and upper limits of each bin, respectively. For the correlations with the zlow bin, this implies a cumulative signal-to-noise ratio for $100 < \ell < 4600$ between 0.1 and 0.3. This contribution becomes larger for the high-redshift source bin (zhigh), from $\mysim 5\%$ to $\mysim 20\%$ of the GGL signal, while the shot and shape noise is of a similar scale. Hence, the cumulative SNR($100 < \ell < 4600$) = 0.2 for the correlation between the zhigh and the first KiDS redshift bin, while the cumulative SNR($100 < \ell < 4600$) = 1.1 between the zhigh and the fifth KiDS bin. At the same time, these $\alpha$ values lead to a maximal contribution of the magnification bias to the clustering signal of $\mysim0.6\%$ \citep{joachimi2020kids}. Even though we are assuming the area of the full 1350 deg$^2$ KiDS footprint, these contributions to the GGL signal by magnification are large enough to prompt the consideration through modelling in the analysis of this syste\-matic in the KiDS-1000+BOSS analysis outlined in \citet{joachimi2020kids}. Nonetheless, since the ana\-lysis shown here already provides an accurate estimate for the magnitude of the magnification bias, the contribution to the GGL signal in each bin can simply be fixed and added to the overall GGL angular power spectrum without the need to add any more free parameters in the astrophysical models.
    
    We note the oscillations at low $\ell$ for some of the zhigh correlations in figure~\ref{fig:kids_ggl_forecasts}. These originate from fluctuations from a power law of $<1\%$ in $C_{\rm{gG}}(\ell)$ for $100 <\ell < 500$. They can be attributed to baryonic acoustic oscillations, BAO, as their amplitude decreases with $\omega_{\rm{b}}$. In figure~\ref{fig:kids_ggl_forecasts} and subsequent forecasts discussed sections~\ref{forecasts:hsc} and \ref{forecasts:Euclid}, the fluctuations in $C_{\rm{mG}}(\ell)/C_{\rm{gG}}(\ell)$ at low $\ell$ appear to be increased to amplitudes $>1\%$ from the mean.  This is caused by the non-BAO signal in $C_{\rm{gG}}(\ell)$ being approximately proportional to $C_{\rm{mG}}(\ell)$ at low $\ell$. Hence, after taking their ratio, the only signal that does not approximately cancel is the BAO signal in $C_{\rm{gG}}(\ell)$. In any case, the variations in in $C_{\rm{gG}}(\ell)$ are well below the uncertainties over that range (which are typically $\gg 2\%$), so they would be undetectable for now.
    
\subsection{HSC Wide + BOSS forecasts}\label{forecasts:hsc}
\begin{figure*}
       \centering
       \includegraphics[width=16cm]{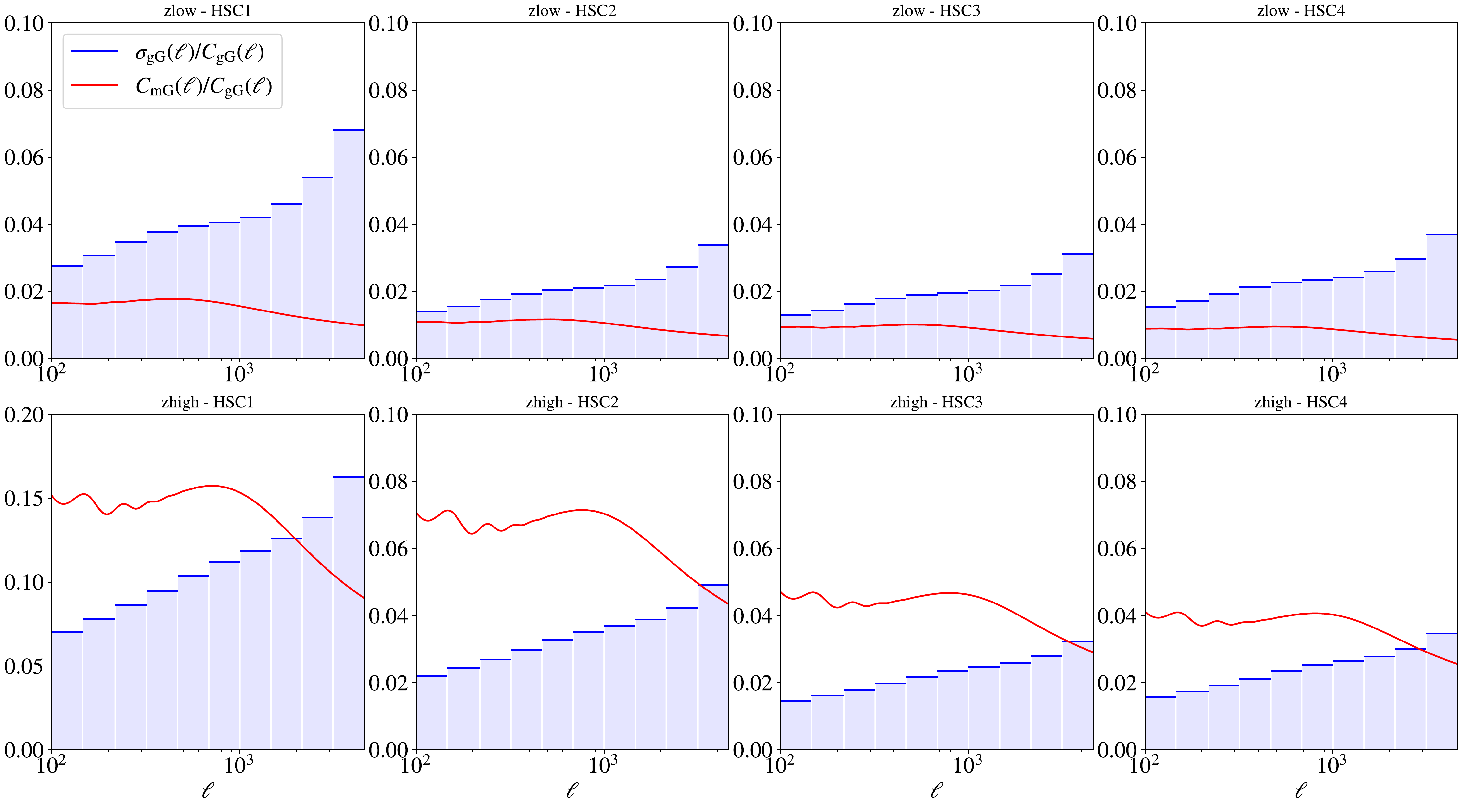}
      \caption{Same as figure~\ref{fig:kids_ggl_forecasts}, but for HSC source bins as defined in table~\ref{table:forecasts} and figure~\ref{fig:all_nofz}.}
    \label{fig:hsc_ggl_forecasts}
   \end{figure*}

    We repeat the analysis for section~\ref{forecasts:kids}, considering the HSC Wide source bins. figure~\ref{fig:hsc_ggl_forecasts} shows the ratio between $C^{\rm{(i, j)}}_{\rm{mG}}(\ell)$ and  $C^{\rm{(i, j)}}_{\rm{gG}}(\ell)$ together with the relative uncertainty in the GGL signal for each $\ell$ bin assuming a full footprint area of 1400 deg$^2$ \citep{aihara2018first} as well as the galaxy sample properties shown in Table~\ref{table:forecasts}. Similar to KiDS, we find that the magnification-shear signal only contributes about $\mysim2\%$ to the GGL signal correlated with the zlow lens bin (giving a cumulative SNR within $100 < \ell < 4600$ between 0.4 and 0.5). In correlations with the zhigh lens bin, the contribution of the magnification-shear signal is larger and between $\mysim 5\%$ and $\mysim 20\%$ which is considerable above the shape and shot noise (with a cumulative SNR within $100 < \ell < 4600$ between 1.3 and 2.0). It is significant enough to give grounds for the consideration of this systematic during future GGL analyses which cross correlate the HSC Wide sample with the BOSS DR12 or a similarly selected lens sample.
    
\subsection{\textit{Euclid}-like survey + DESI-like survey forecasts} \label{forecasts:Euclid}
     We produce forecasts for a GGL analysis with Stage-IV \citep{albrecht2006report}, assuming lens and source samples akin to DESI \citep{aghamousa2016desi} and \textit{Euclid} \citep{laureijs2011euclid}, respectively. We repeat the analysis shown in section~\ref{forecasts:kids} and \ref{forecasts:hsc} for the \textit{Euclid}-like source bins described in table~\ref{table:forecasts} and in figure~\ref{fig:all_nofz}. We consider a footprint overlap between our source and lens sample of 6000 deg$^2$, which is roughly the expected overlap between \textit{Euclid} and DESI \citep{levi2013desi, aghamousa2016desi}. Therefore, the fictitious BOSS/DESI-like galaxy sample we are considering here has all the properties of the BOSS lens sample, but has the planned DESI footprint. Although DESI will probe higher redshifts and fainter galaxies than BOSS, it will be similar to BOSS in that it will not be a purely flux-limited survey. Targets in DESI are selected using a combination of different band magnitudes depending on the galaxy type and redshift range which is being observed (for more details see \citealt{aghamousa2016desi}). For this reason, the magnification bias in the DESI sample cannot be modelled analytically either, warranting an analysis similar to the one discussed here. The \textit{Euclid}-like source sample used in this work is designed to be split into the same redshift bins as suggested by \textit{Euclid} collaboration forecast choices \citep{blanchard2019euclid}. In addition, within each bin, the median redshift is chosen to be in agreement with the one expected for the \textit{Euclid} sources.\\
     
\begin{figure*}
       \centering
       \includegraphics[width=17cm]{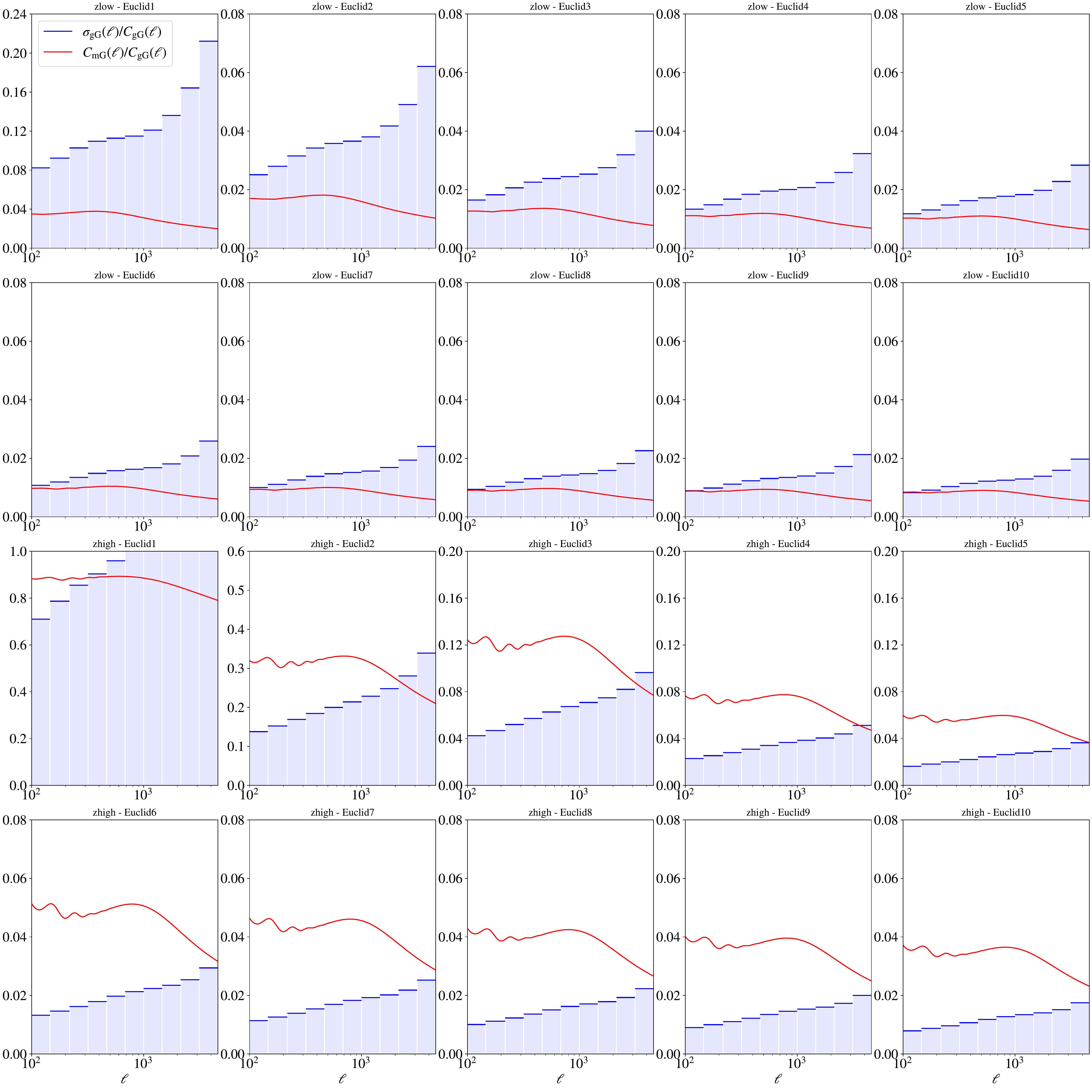}
      \caption{Same as figure~\ref{fig:kids_ggl_forecasts}, but for \textit{Euclid}-like source bins as defined in table~\ref{table:forecasts} and figure~\ref{fig:all_nofz}.}
    \label{fig:Euclid_ggl_forecasts}
   \end{figure*}

     Considering 6 logarithmically spaced $\ell$ bins per dex (as in the previous sections), we obtain the magnification-shear signal forecasts shown in figure~\ref{fig:Euclid_ggl_forecasts}. We see that the magnification-shear signal constitutes a considerable systematic when correlating with the zlow bin, since the observed cumulative SNR on scales within $100 < \ell < 4600$ is between 0.3 and 0.7. The magnification bias signal becomes strong enough for correlations with zhigh, it would be a detectable signal (with the cumulative SNR  within $100 < \ell < 4600$ ranging from 1.5 when correlating zhigh and Euclid1 to 2.8 when correlating zhigh and Euclid10). This might require any future GGL analysis of \textit{Euclid}+BOSS or \textit{Euclid}+DESI data to allow for the $\alpha$ parameters to freely vary as a nuisance parameter in order to properly account for this systematic. The method outlined in this paper could be used to set informative priors on the $\alpha$ values within each lens bin.
     
\section{Conclusions} \label{conclusion}

In this paper, we have introduced a novel method to estimate the effective luminosity function slope, $\alpha$, of galaxy samples which have been defined with a complex selection function that is not simply flux/magnitude-limited. The method calibrates the $\alpha$ estimates from observables with accurate cosmological simulations with the same sample selection. This expands upon previous work where the flux magnification was only measured for flux-limited cases or found to be inaccurate in non-flux-limited cases \citep{hildebrandt2016observational}.\\

The new method determines the underlying slope of the lumino\-sity function of the simulated galaxy sample ($\alpha_{\kappa}$) from unobservable properties such as the convergence, $\kappa$, and the unlensed galaxy position. It then finds the magnitude range relative to the magnitude limit over which the resulting $\alpha_{\rm{obs}}$ as calculated from the observable differential galaxy count distribution, $n(m)$, best agrees with $\alpha_{\kappa}$. Finally, the same relative magnitude range is used to determine $\alpha_{\rm{obs}}$ from the observed galaxy sample.\\ 

A few things should be considered when employing this method. We find that the magnitude ranges up to the effective magnitude limit that are determined to be optimal from the simulations in order to calibrate $\alpha_{\rm{obs}}$ are only valid for a given redshift range, a given sample selection function and a given galaxy sample for which weak lensing simulations are available. Thus, it is important to note that this method cannot be generalised trivially, as it requires the availability of accurate cosmological simulations to assure consistency between the two independent $\alpha$ estimates, $\overline{\alpha}_{\rm{obs}}$ and $\overline{\alpha}_{\kappa}$. Nonetheless, when simulations are available, it provides a robust estimate of the scale of the magnification bias for non-flux-limited surveys such as BOSS.\\

Applying our calibration method to the BOSS DR12 sample split into two redshift bins, we find that $\overline{\alpha}_{\rm{obs}} = 1.93 \pm 0.05$ for $0.2< z \leq0.5$ and $\overline{\alpha}_{\rm{obs}} = 2.62 \pm 0.28$ for $0.5 < z \leq 0.75$ leading to a contribution to the galaxy-galaxy lensing signal of up to $\mysim2\%$ for KiDS-1000 and HSC Wide sources correlated with the $0.2<z \leq 0.5$ lens bin. Although the contribution can go up to $\mysim 20\%$ when correlating KiDS-1000 and HSC Wide sources with the $0.5 < z \leq 0.75$ BOSS lens bin, the magnification-shear signal can go above the noise with a cumulative SNR going up to 1.1 and 2.0 for KiDS-1000 and HSC Wide, respectively. Hence, both for KiDS-1000 and HSC Wide, the magnification-shear signal appears to be dominant enough to warrant the modelling of this systematic in future GGL analyses involving BOSS lenses, as was already done in the recent KiDS-1000 analysis \citep{joachimi2020kids, heymans2020kids}. This necessity becomes even more evident in the forecasts for a GGL analysis of \textit{Euclid}-like sources with DESI-like lenses. In this case, the magnification-shear signal is either a considerable systematic when correlating with the zlow bin (with a cumulative SNR of around 0.5), or it even becomes a detectable signal when correlated the source bins with zhigh giving cumulative SNRs around 2 which can go up to 2.8. This might require any future GGL analysis incorporating \textit{Euclid} and any highly selected lens sample (e.g. DESI or BOSS) to allow for the effective luminosity function slope ($\alpha$) of each lens sample to vary freely within the model using informative priors based on an analysis similar to the one conducted in this paper. These results are in line with \cite{duncan2014complementarity}  as well as the recent findings from \cite{mahonyprep} where it was determined that the inclusion of the magnification bias in the modelling for surveys such as the next generation of surveys is necessary to accurately infer cosmological parameters. \\

We expect similar conclusions for other surveys. It might be desirable to estimate the magnification bias using the methodo\-logy outlined in this paper in clustering and GGL ana\-lyses based on DES \textsc{redMaGiC} lens galaxies such as the ones described in \cite{clampitt2017galaxy}, \cite{elvin2018dark} and \cite{prat2018dark}, since it also follows a complex selection function \citep{rozo2016redmagic}. The SNR should be comparable to HSC and KiDS, so the magnification bias will not have to be included as a free parameter. On the other hand, for surveys such as LSST \citep{ivezic2008lsst, abell2009lsst} and the Nancy Grace Roman Space Telescope (formerly known as WFIRST, \citealt{spergel2015wide}), it may become necessary to make the $\alpha$ of the lens galaxy samples a nuisance parameter in any clustering or GGL analysis, as we suggest for a \textit{Euclid}+DESI-like analysis.
\newpage
\section*{Acknowledgements}

We thank our referee for a constructive and comprehensive report. We would also like to thank Chieh-An Lin for helpful discussions. MWK thanks the Science and Technology Facilities Council for support in the form of a PhD Studentship. We acknowledge support from the European Research Council under grant numbers 647112 (CH, MA, TT) and 770935 (HH, AW,JvdB), as well as the Deutsche Forschungsgemeinschaft (HH, Heisenberg grant Hi 1495/5-1). CH and SU also acknowledge support from the Max Planck Society and the Alexander von Humboldt Foundation in the framework of the Max Planck-Humboldt Research Award endowed by the Federal Ministry of Education and Research. 

The MICE simulations have been developed at the MareNostrum supercomputer (BSC-CNS) thanks  to grants AECT-2006-2-0011 through AECT-2015-1-0013.
This work has made use of CosmoHub \citep{Carretero17}. CosmoHub has been developed by the Port d'Informació Científica (PIC), maintained through a collaboration of the Institut de Física d'Altes Energies (IFAE) and the Centro de Investigaciones Energéticas, Medioambientales y Tecnológicas (CIEMAT), and was partially funded by the "Plan Estatal de Investigación Científica y Técnica y de Innovación" program of the Spanish government.
\newpage
\section*{Data Availability}
The SDSS-III BOSS DR12 data \citep{eisenstein2011sdss, dawson2012baryon} underlying this article is available at \url{https://www.sdss.org/dr12/} and the datasets generated by the MICE2 simulations \citep{Fosalba15a,Fosalba15b,Carretero15,Crocce15,Hoffmann15} used in this work are accessible on CosmoHub (\citealt{Carretero17}, \url{https://cosmohub.pic.es/home}). The code used for the ana\-lysis presented in this paper can be found in the \textsc{MagBEt} GitHub repository (\url{https://github.com/mwiet/MAGBET}).



\bibliographystyle{mnras}
\bibliography{magbias}

\appendix

\section{Flux-limited case} \label{flux-lim}
    As discussed in section~\ref{method:data}, we conduct a sanity check of our method by comparing the effective $\overline{\alpha}_{\kappa}$ to $\overline{\alpha}_{\rm{obs}}$ for each redshift bin given a simulated magnitude-limited ($i<20.2$) galaxy population spanning the whole sky. This sample is also based on MICE2 simulations. We estimate $\overline{\alpha}_{\kappa}$ from the known matter convergence $\kappa$ and the relative difference between the lensed and unlensed cumulative galaxy number counts finding that $\overline{\alpha}^{\rm{zlow}}_{\kappa} = 0.97 \pm 0.13$ in the zlow bin ($0.2 < z \leq 0.5$) and that $\overline{\alpha}^{\rm{zhigh}}_{\kappa} = 3.15 \pm 0.10$ in the zhigh bin ($0.5 < z \leq 0.75$).\\
    
      \begin{figure}
      \centering
       \includegraphics[width=7.5cm]{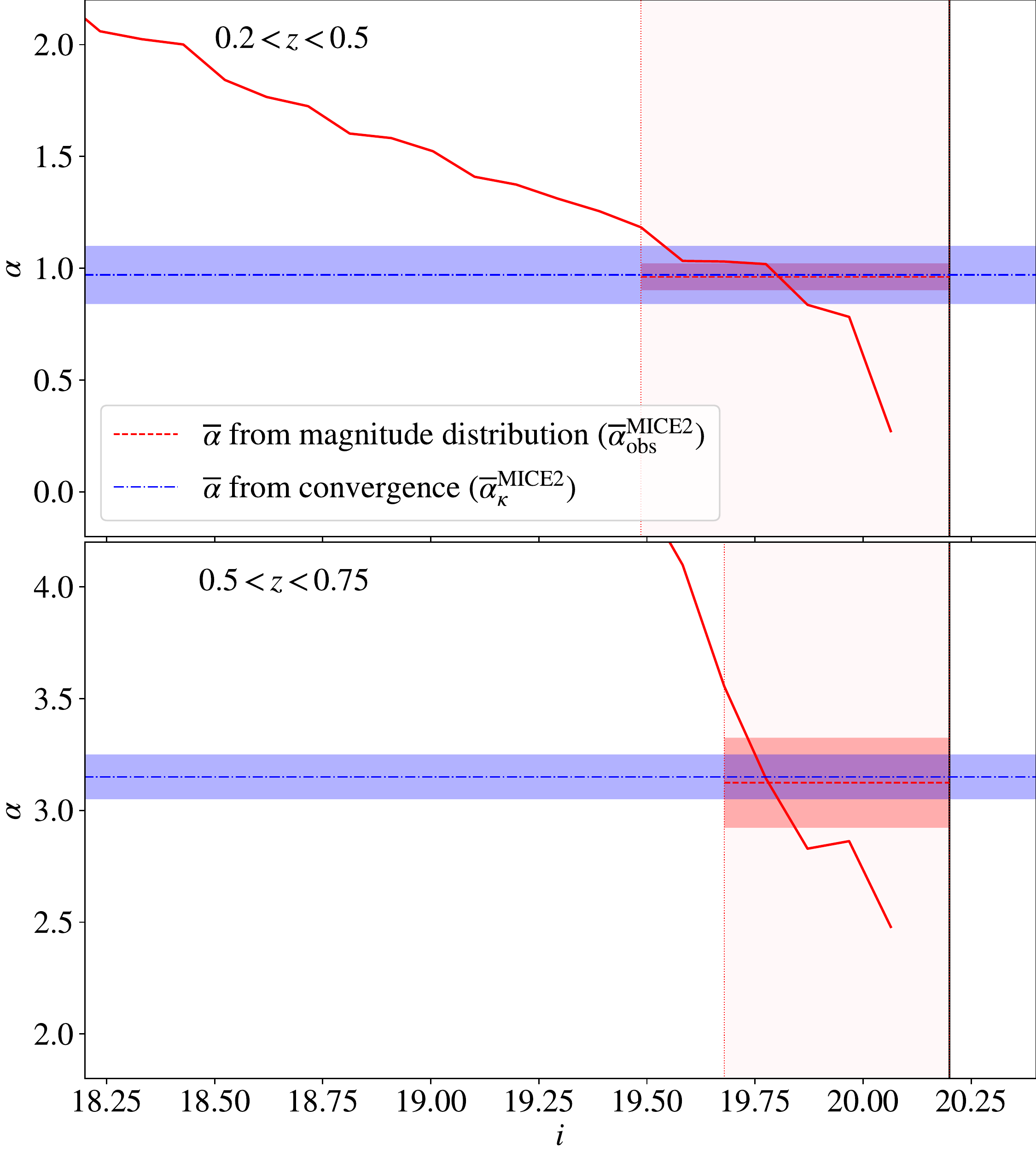}
          \caption{The slope of the luminosity function, $\alpha$, as a function of the $i$-band magnitude, $i$, for the magnitude-limited case. Two redshift bins are considered: $0.2 < z \leq 0.5$ (top) and $0.5 < z \leq 0.75$ (bottom). The vertical black line marks the magnitude limit at $i = 20.2$ and the dashed red vertical lines mark the upper and lower bounds of the highlighted magnitude range which was used to determine $\overline{\alpha}_{\rm{obs}}^{\rm{MICE2}}$. The
          dashed red horizontal line marks the $\overline{\alpha}_{\rm{obs}}^{\rm{MICE2}}$ estimate and the blue dot-dashed horizontal line marks the effective $\overline{\alpha}^{\rm{MICE2}}_{\kappa}$ determined from the weak lensing convergence with equation~(\ref{eqn:reldiff}) and used to calibrate $\overline{\alpha}_{\rm{obs}}^{\rm{MICE2}}$.}
         \label{fig:flux-lim-alpha-hist}
   \end{figure}
   \begin{figure}
       \centering
       \includegraphics[width=7.5cm]{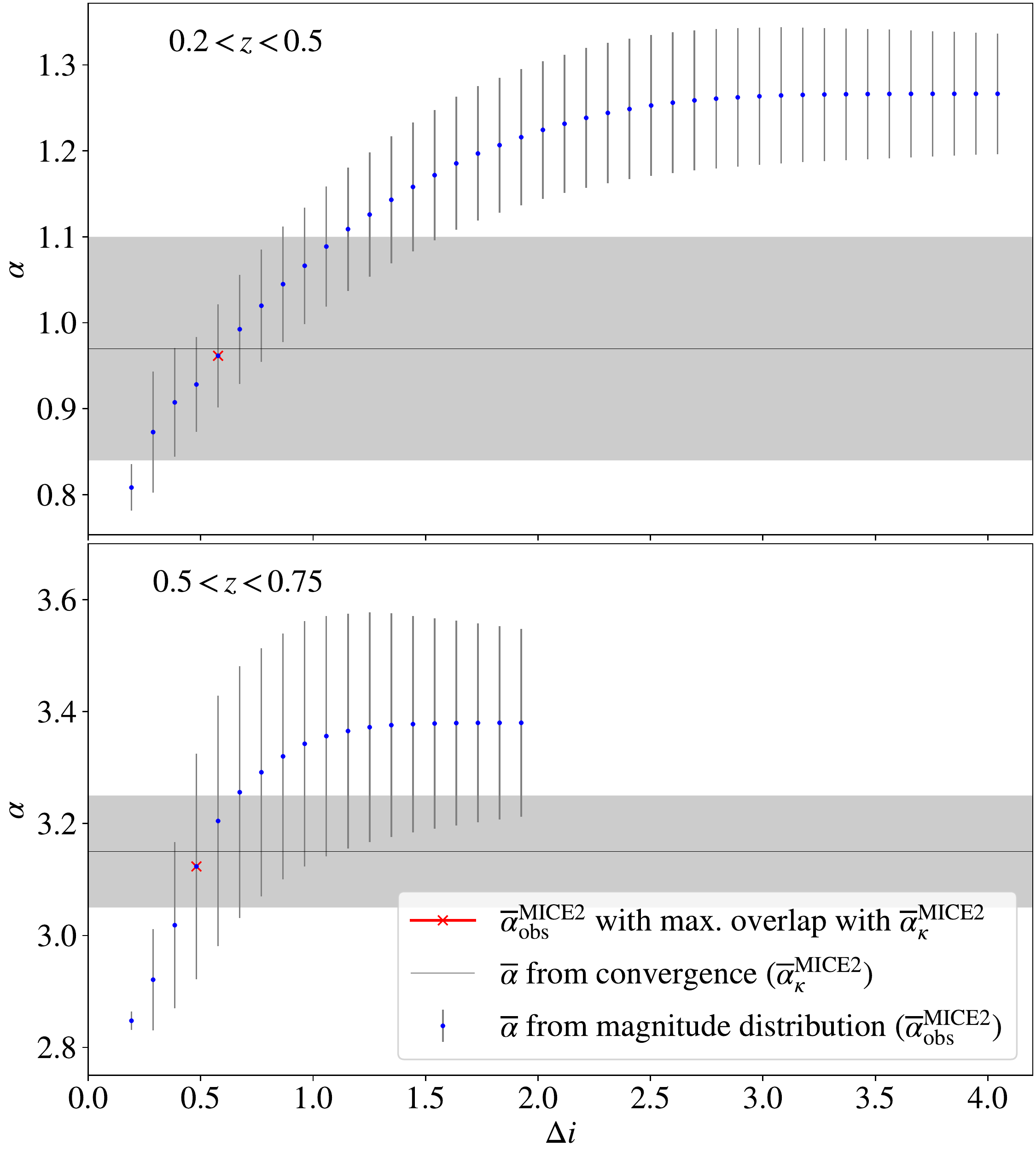}
          \caption{$\overline{\alpha}_{\rm{obs}}$ estimates from the MICE2 simulations for the magnitude-limited case ($i < 20.2$) over $i$-band magnitude ranges below the turn-off magnitude ($\Delta i$) considered to calculate the weighted average. Two redshift bins are considered: $0.2 < z \leq 0.5$ (top) and $0.5 < z \leq 0.75$ (bottom). The red cross marks the $\overline{\alpha}_{\rm{obs}}$ estimate which overlaps the most with the $\overline{\alpha}_{\kappa}$ estimate from the weak lensing convergence (black line).}
         \label{fig:flux-lim-alpha-varying}
   \end{figure}
    
    The $\overline{\alpha}_{\kappa}$ values are compared to $\overline{\alpha}_{\rm{obs}}$ in figure~\ref{fig:flux-lim-alpha-hist}. We find that in zlow, $\overline{\alpha}_{\rm{obs}}$ optimally overlaps with the $\overline{\alpha}_{\kappa}$ estimate from the convergence when taking the weighted mean of $\alpha_{\rm{obs}}(m)$ over a magnitude range of $\Delta i = 0.67$ below the effective magnitude limit; giving $\overline{\alpha}_{\rm{obs}}^{\rm{zlow}} = 0.96 \pm 0.06$. This agrees well with the $\overline{\alpha}_{\kappa}$ of galaxies in zlow ($\overline{\alpha}^{\rm{zlow}}_{\kappa} = 0.97 \pm 0.13$). The agreement is similarly good in the zhigh bin where the optimal $\overline{\alpha}_{\rm{obs}}$ is computed over a $\Delta i = 0.48$ and found to be $\overline{\alpha}_{\rm{obs}}^{\rm{zhigh}} = 3.12 \pm 0.20$. The excellent agreement between $\overline{\alpha}_{\kappa}$ and when evaluating near the faint end of the sample ($\Delta i < 0.7$) reinforces that equation~(\ref{eqn:reldiff}) and equation~(\ref{eqn:alpha_obs}) indeed describe the same $\alpha$; at least for the flux-limited. Such a good agreement is not really surprising, since the underlying assumptions leading to equation~(\ref{eqn:reldiff}) ($|\kappa| \ll 1; |\gamma|\ll 1$) are ingrained in the way the MICE2 simulations determine the magnified magnitude and position of galaxy \citep{Fosalba15a}. Nonetheless, it still provides a check which allows us to understand how the method estimates $\alpha$ in the absence of a complex selection function .\\
    
    In addition, when looking at figure~\ref{fig:flux-lim-alpha-varying}, one finds that for zlow the $\overline{\alpha}_{\rm{obs}}$ estimates are accurate over a large domain of magnitude ranges (being less than $1\sigma$ apart when considering $\Delta i$ anywhere between 0 and $\mysim2$). This confirms that a power law is a good approximation for the luminosity function over a large magnitude range near the faint end of the distribution which implies that, in a magnitude-limited survey, $\overline{\alpha}_{\rm{obs}}$ estimates are robust and accurate even after substantial changes in the magnitude range considered. We find similarly good agreement between the $\overline{\alpha}_{\kappa}$ and $\overline{\alpha}_{\rm{obs}}$ in the zhigh bin where $\overline{\alpha}_{\rm{obs}}^{\rm{zhigh}} = 3.12 \pm 0.2$, while figure~\ref{fig:flux-lim-alpha-varying} shows that this estimate is robust at high redshifts.\\
    
    Despite the consistency between $\overline{\alpha}_{\kappa}$ and $\overline{\alpha}_{\rm{obs}}$ and the robustness of the estimate to small changes in the calibration magnitude range $\Delta i$, it is surprising to see such a drastic increase in $\overline{\alpha}_{\rm{obs}}$ between zlow and zhigh. This seems to be a consequence of the magnitude limit at $i = 20.2$ being low enough to exclude a substantial fraction of faint galaxies at high redshifts, such that the power law in flux assumed in equation~(\ref{eqn:n0}) no longer applies. If we consider the luminosity function of the galaxies as a Schechter function \citep{schechter1976analytic}, such a selection of bright galaxies would lead to a dominant exponential term in the Schechter function which leads to overestimates of $\alpha$. In general, this is not of much concern, since most magnitude limited surveys operate within a regime where the power law approximation holds.\\
    
\bsp	
\label{lastpage}
\end{document}